\documentclass[10pt,journal,letterpaper,compsoc]{IEEEtran}
\makeatletter
\def\ps@headings{%
\def\@oddhead{\mbox{}\scriptsize\rightmark \hfil \thepage}%
\def\@evenhead{\scriptsize\thepage \hfil\leftmark\mbox{}}%
\def\@oddfoot{}%
\def\@evenfoot{}}
\makeatother 
\usepackage{amsfonts}
\usepackage{bbm}
\usepackage{mathrsfs}
\usepackage{cite}
\usepackage{times}
\usepackage{graphicx}
\usepackage{amsfonts}
\usepackage{flushend}
\usepackage{algorithmic}
\usepackage[ruled]{algorithm}
\usepackage{cite}
\usepackage{times}
\usepackage{amsmath}
\usepackage{graphicx}
\usepackage{epsfig}
\usepackage{amsmath}
\usepackage{latexsym}
\usepackage{amsfonts}
\usepackage{amssymb}
\usepackage{paralist}
\usepackage{textcomp}
\usepackage{xspace}
\usepackage{mathrsfs}
\usepackage{amssymb}
\usepackage{color}
\usepackage{algorithmic}
\usepackage[ruled]{algorithm}
\usepackage{flushend}
\usepackage{cite}
\usepackage{times}
\usepackage{graphicx}
\usepackage{amsfonts}
\usepackage{flushend}
\usepackage{cite}
\usepackage{times}
\usepackage{amsmath}
\usepackage{graphicx}
\usepackage{graphics}
\usepackage{epsfig}
\usepackage{amsmath}
\usepackage{latexsym}
\usepackage{amsfonts}
\usepackage{amssymb}
\usepackage{paralist}
\usepackage{xspace}
\usepackage{mathrsfs}
\usepackage{amssymb}
\usepackage{color}
\usepackage{epstopdf}
\usepackage{algorithmic}
\usepackage{algorithm}
\usepackage{subfigure}
\usepackage{flushend}

\usepackage{booktabs}
\usepackage{tabularx}
\usepackage{subfigure}
\usepackage{subfig}
\usepackage{epstopdf}

\allowdisplaybreaks 

\renewcommand{\paragraph}[1]{\smallskip \noindent {\textsc{#1}}}

\newcommand{\mycut}[1]{{}}
\DeclareGraphicsExtensions{.eps,.ps,.pdf}

\begin{document}


\title{Artificial Impostors for \\  Location Privacy Preservation}

\author{Cheng~Wang,~\IEEEmembership{Senior Member,~IEEE},
and Zhiyang Xie
   \IEEEcompsocitemizethanks{
   \IEEEcompsocthanksitem  Cheng Wang and Zhiyang Xie are with the Department of Computer Science and Engineering, Tongji University, and with the Key Laboratory of Embedded System and Service Computing, Ministry of Education, China. (E-mail: cwang@tongji.edu.cn, 121xzy@tongji.edu.cn)}
}

\IEEEcompsoctitleabstractindextext{

\begin{abstract}
The progress of location-based services has led to severe concerns about location privacy leakage.
However, existing methods are still incompetent for effective and efficient location privacy preservation (LPP).
They are often vulnerable under the identification attack with side information,
or hard to be implemented due to the high computational complexity.
In this paper, we pursue the high protection efficacy and low computational complexity simultaneously.
We propose a \emph{scalable} LPP method based on the paradigm of counterfeiting locations.
To make fake locations  extremely plausible,
we forge them by synthesizing \emph{artificial impostors} (AIs).
The so-called AIs refer to the synthesized traces that have similar semantic
features to the actual traces and do \emph{not} contain any target location.
We devise two dedicated techniques:
the \emph{sampling-based synthesis method} and \emph{population-level semantic model}.  They respectively play the significant roles in two critical steps of synthesizing AIs.
We conduct the experiments on real datasets in two cities (Shanghai of China and Asturias of Spain) to validate the high efficacy and scalability of the proposed method.
In these two datasets, the experimental results show that our method achieves the preservation efficacy of $97.68\%$ and $96.24\%$, and the time spent for building generators is only $230.47$ and $215.92$ seconds, respectively.
This study would give the research community new insights into improving the practicality of the state-of-the-art LPP paradigm via counterfeiting locations.
\end{abstract}
}

\maketitle

\IEEEdisplaynotcompsoctitleabstractindextext
\IEEEpeerreviewmaketitle
\section{Introduction}

With the proliferation of smart mobile terminals, e.g., smartphones and vehicle-mounted communication devices, and the location-based services (LBS) over them, the privacy issues of locations have attracted much attention \cite{Eckhoff2014Driving}.
Location privacy usually refers to the demand to prevent other parties from learning one's current or past locations \cite{beresford2003location}.
The obfuscation mechanism is widely utilized to protect the location privacy, which transforms an actual location to a set of locations on the trusted proxy. Then the providers of LBS answer queries to all locations in this set, and the trusted proxy returns the useful information to the user.

Some obfuscation methods generate a location set by blending the actual location with locations of other clients
\cite{beresford2003location,lee2011protecting,yao2012clustering}. However, these methods possibly leak the privacy of other real clients.
Some methods replace user's location with nearby points of interest  \cite{khoshgozaran2007blind,yiu2008spacetwist}. To some extent, the substitutes might reveal the region where the user is.
Other methods generate fake locations based on different moving patterns. They mix an actual location with fake ones in a set \cite{synthisiziengsandp2016,krumm2009realistic,chow2009faking,you2007protecting}.
Unfortunately, these protection methods are vulnerable to inference attacks or suffer from substantial time costs.
The adversary can easily identify the fake ones with some side information.
Taking this condition as an instance:
The fake location is in a supermarket at 2:00 am. According to common sense, the supermarket is closed at this time.
This kind of attack uncovers the fake ones through the semantic relations between the locations and time or activities.
These fake locations are too \emph{implausible} to hide the real one.
In other words, the goal of this kind of obfuscation mechanism is to make the fake ones \emph{plausible} in the LBS server side.

In this paper, we aim at designing a scalable and high-quality location privacy preserving method.
The preservation quality/efficacy of a model is defined as the attacker's probability of error to infer the real trace \cite{shokri2011quantifying}.
To achieve the high efficacy of privacy protection, we generate fake locations by synthesizing artificial impostors.
The so-called impostors refer to the synthesized traces which have similar semantic
features to the actual traces, and do not contain real locations of queries.
It is worth mentioning that a location refers to a region in our method for the reason of location cloaking.
To make impostor traces plausible, we utilize visiting patterns of regions and mobility patterns of users.
Visiting pattern refers to the visitors' temporal distributions of regions.
Mobility pattern refers to the transition probabilities among regions and the runtime of passing through regions of users.
There are two steps in our method:
The first is to extract regional semantic features based on the visiting pattern (offline generator).
The second is to generate impostor traces based on the mobility pattern (online generator).
Accordingly, we devise two critical techniques in two steps to achieve a high efficacy with a low computational complexity:

(1)  \emph{A Sampling-Based Synthesis Method}.
Users' trajectories are composed of their visited locations.
A straightforward method for synthesizing an impostor trace is to forge fake locations corresponding to all the visited locations.
However, this complicated method \emph{cannot} improve the preservation efficacy \emph{considerably}.
The requirement of too many simultaneously plausible fake locations possibly limits the number of candidate plausible traces, then let down the preservation efficacy.
When generating a plausible but \emph{fake} trace for a user,
it is convincingly unnecessary to utilize all the locations in his/her trajectory.
Our solution is to sample visited locations directly.
The challenge here is how to keep the plausibility of artificial impostor traces based on the sampled locations.
To this end, we only choose some special pause points (not the locations just passed through) as milestones along the trip.
 We call them \emph{stations}. They are semantic regions where people want to go instead of simple sampling locations in the trace.
For example, Alice drives home after work and wants to buy a hamburger in a drive-through restaurant.
There are three stations along her trace: Her company, the restaurant, and her home.
Except for these three stations, the locations she passed by are relatively meaningless in the privacy sense.
This kind of sampling method is \emph{process-independent}, which means we utilize stations to synthesize fake traces, but not all the locations in the trace.
Based on stations, we synthesize an impostor trace by the following procedures:
Replacing stations with fake ones (semantically similar locations), and then complementing plausible paths among stations by finding the $k$th most possible traces according to the transition probabilities among regions.

(2)  \emph{A Population-Level Semantic Model}.
Station refers to the record with the location and time.
To make the generated fake station resemble the real one, we let the fake and real stations have similar semantic regional features.
In the process of extracting the regional semantic features, a critical step is the location clustering.
Each cluster of locations has a semantic feature.
To dig the semantic similarity among locations concerning visited patterns,
a useful approach is to analyze the transfer probability among visited locations along traces of different users.
Unsurprisingly, such an \emph{individual-level} approach is so careful that it can complete semantic clustering with impressively high efficacy.
However, a substantial computational cost of this extremely carefulness is indeed daunting, though the achieved high efficacy is indeed enviable.
In this work,
under the precondition of high efficacy,
we strive for a low computational cost by devising a \emph{population-level} model to extract features and cluster locations.
We divide the city map into different regions, and break up all trajectories into independent locations.
After that, we extract stations of users' traces and model the geographical distribution of these stations in different regions.
The feasibility of our population-level model depends on the fact that the essential semantic feature of a region can be judged from the mobility of crowds.
Consider a simple situation where most people leave home to work in the morning, and come back home at night. It is evident that the out-stream is abundant in the morning and the in-stream is abundant in the night of uptown, whereas the opposite is the statistical stream of people in a workplace.
Hence, based on this geographical distribution model, we can compute
the semantic similarity of different regions, and then aggregate all regions into clusters with different latent semantic features.

With the cooperation of these two techniques, our method successfully achieves high computational
efficiency with conspicuous preservation efficacy. Intuitively, the sampling-based synthesis model
can to some extent subserve the improvement of preservation efficacy
despite the reduction due to the coarse clustering by the population-level model.

In the experiment, we compare our method with four representative ones on the different scales of maps in two cities (Shanghai of China and Asturias of Spain). We validate the performance by the state-of-the-art location inference attack \cite{shokri2011quantifying}. The primary experimental results in these two datasets (Shanghai's and Asturias's, respectively) can be summarized as follows:

$\bullet$  In the small-scale map ($12\mbox{km}\times9\mbox{km}$), the preservation efficacy of our method is slightly lower than SG-Lppm \cite{synthisiziengsandp2016}, the state-of-the-art method of location privacy preservation, but is significantly higher than other three methods.
Moreover, our method has a significant advantage in the time consumed by generating a fake trace online over SG-Lppm.
Specifically, our method achieves the efficacy of
$97.95\%$ and $ 95.91\%$, compared with $98.33\%$ and $97.59\%$  by SG-Lppm in two datasets, respectively.
While other methods achieve the efficacy which is less than $80\%$.
To generate one fake trace online,
it takes our method only $19.76$ms and $17.89$ms, compared with $72.31$s and $74.03$s by SG-Lppm in two datasets, respectively.

$\bullet$ In the large-scale map ($42\mbox{km}\times35\mbox{km}$) of two cities, we build the impostor generator offline in just $230.47$ and $215.92$ seconds in two datasets, and we can achieve the efficacy of $97.68\%$ and $96.24\%$, respectively. However, SG-Lppm has been executed for two weeks without any output, due to its high time complexity.

The rest of this paper is organized as follows:
In Section \ref{sec-relatedwork}, we provide the related work. We give an overview of our proposed scheme in Section \ref{sec-overview}, and
describe two main steps of the scheme in Section \ref{sec-featurepart1} and Section \ref{sec-tracespart2}, respectively.
We provide the validation of  preservation efficacy and the evaluation of scalability of our methods in  Section \ref{sec-evalu}.
Finally, we draw a conclusion in Section \ref{sec-conclu}.

\section{Related Work}\label{sec-relatedwork}

There has already existed plenty of remarkable studies on location privacy preservation
\cite{synthisiziengsandp2016,Cheng2006Preserving,Ardagna2010An,DBLP:conf/mobihoc/DouZLYGGRL16,DBLP:journals/tdsc/LiuWCRZD14,DBLP:journals/tvt/LiuZZZL17}.
According to the application scenarios,  the privacy preservation methods can be widely classified into three types  \cite{Miguel2012Geo}.

The first type of techniques is proposed to protect the privacy of location during transmission.
Khoshgozaran et al. \cite{khoshgozaran2007blind} defined a Hilbert ordering with a key, and use it to encrypt the database and the queries.
Ashouri et al. \cite{Ashouri2012GLP} proposed the GLP protocol to address the location privacy problem for a group of users.
These methods are based on cryptographic methods.
The adversary can not decode information which is transmitted between users and LBS providers.

The second type of techniques is devised to protect the privacy of the statistic database of location data. The state-of-the-art technique is differential privacy \cite{Miguel2012Geo,Zhang2014Differentially,Enabling2018}. Its goal is to protect an individual's data while publishing aggregate information about the database.
Andres et al. \cite{Miguel2012Geo} proposed to achieve Geo-Indistinguishability by adding noise drawn from a Laplace distribution.
Zhang et al. \cite{Enabling2018} proposed a new probabilistic differential Privacy-preserving location recommendation framework. They used it to achieve the trade-off between high recommendation accuracy and strict location privacy for check-in dataset.

The third type of techniques is utilized to protect the privacy of individual-specific location data. This kind of techniques is based on obfuscation methods.
They transform an actual location to a set of locations, which keeps the utility of services and privacy of location.
This is the focus of this work.
According to the different ways of generating location sets, we classify these methods into $3$ categories as follows:

(1)  Anonymizing the actual location in a set of locations of other users.
This anonymization mechanism is a function to hide the user in the class with $k$ users in
\cite{beresford2003location,Hoh2005Protecting,lee2011protecting,yao2012clustering,Kirkpatrick2012Privacy,hara2014location,Chow2009Casper}.
Hara et al. \cite{hara2014location}, selected the fake traces among others with some strict constraints for a targeted real trace. Although this protection performs well, other's privacy was leaked during this process.
Yao et al. \cite{yao2012clustering} divided the area into clusters (namely $CK$) where each cluster includes $k$ users.
When a client proposes a query to the LBS provider, the boundary of the area he belongs to is sent to the servers.
However, this method limits an adversary's probability to infer the accurate location to $1/k$.
If there are no $k-1$ clients near the user, the $CK$ will be too large to provide satisfactory service. Moreover, this exposes the general location of them.
Lee et al. \cite{lee2011protecting} extracted the  semantic features of regions by the distribution of nearby users' staying duration of locations.
By the semantic features, they computed the cloaking area.
However, the approach is invalid when there are not enough other users around.

(2) Replacing the actual location with a set of other places. This mechanism is a kind of methods to replace the real location with the place near-by (e.g., spatial cloaking)
\cite{Ardagna2010An,Freudiger2013Non,yiu2008spacetwist,khoshgozaran2007blind}.
Dewri et al. \cite{Dewri2010Query} utilized the principle of m-invariance to generate spatial cloaking regions to protect location privacy.
To generate the cloaking regions, they adopted and modified
HilbertCloak algorithm introduced in [16].
Yiu et al. \cite{yiu2008spacetwist} transformed the user's location to intersection or building nearby. However, the distance between the real location and target has an impact on the utility and privacy. If there is no target near the user, the response from LBS provider cannot match the real query accurately, and if the target is close enough to the user,
it may expose his/her location.
The spatial transformation method \cite{khoshgozaran2007blind} uses Hilbert curves to transform users' locations
and sends the transformed location to the LBS server. The disadvantage of this method is that it requires LBS providers to transform all locations data (such as locations of shops).
The maintaining cost of services is noticeable.

(3) Generating a set with the real location  and generated dummy locations in \cite{Bindschaedler2015Privacy,Shokri2015Privacy}.
Generating dummy locations aims to hide the real location among a set of fake locations.
The LBS provider makes the response to all the queries, and TTP filters the information of the user's requirement.
So, this kind of methods can keep the service utility. To improve the effectiveness, dummy locations can be generated based on synthesizing traces.
You et al. \cite{you2007protecting} generated dummy locations based on a random walk.
Chow \cite{chow2009faking} built the path between two random locations on the map.
Xu et al. \cite{Xu2007Location} generated the fake paths based on a simple greedy algorithm and spatial generalization strategy.
To make fake locations more plausible, locations can be classified by semantic features.
The method in \cite{lee2011protecting} generates semantic features by the time durations of places, \cite{yuan2012discovering} by combining human mobility with POIs,  and \cite{do2014places} by computing the frequency of visiting locations.
On this basis, Shorkri et al.\cite{synthisiziengsandp2016} extracted the mobility patterns as common semantic features through matching the locations of every two users in the real mobility datasets, and replaced all locations of a user's real trace with ones in the same semantic cluster to generate fake traces.
This method takes the correlation of sequential locations into account, so that extend generating fake locations to fake traces.
This approach achieves a high efficacy in their paper.
However, while enjoying such efficacy, it has to suffer from a considerable time cost to generate semantic clusters of regions, and more simultaneously plausible fake locations limit the number of candidate plausible traces, then possibly let down the applicability.

We limit the scope of our study into the third type of techniques mentioned above, i.e.,
the preservation methods that are utilized to protect the privacy of individual-specific location data.
This is one of the reasons why there are no comparisons with the methods in different scenarios in our experiments,
such as the differential privacy preservation, despite of their well-recognized high performance in specific application scenarios.

\section{Overview of Proposed Scheme}\label{sec-overview}

\begin{figure}
\centering
\includegraphics[width=0.4\textwidth]{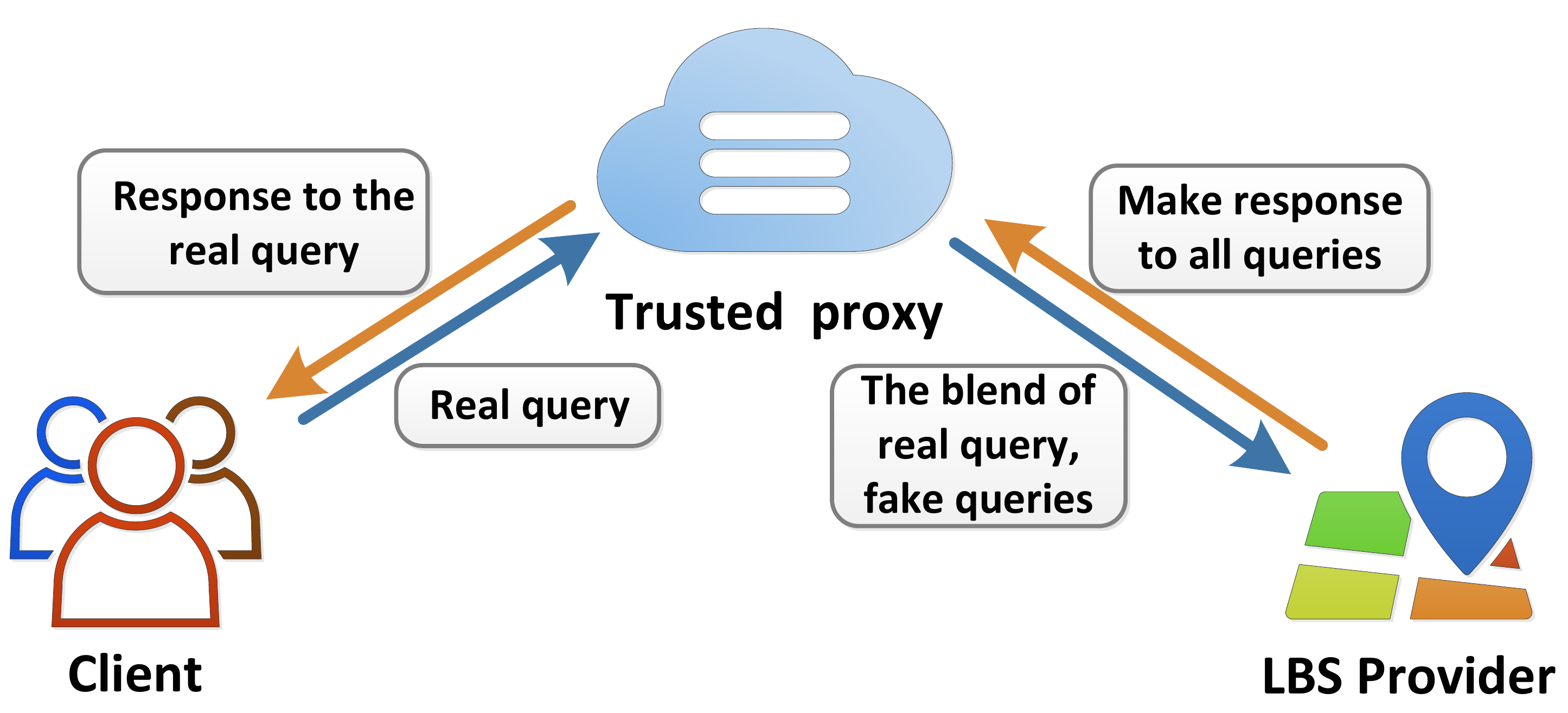}
\caption{When a user proposes a query about the record in the $i$th time interval to the LBS provider (e.g., the nearby restaurants of the current location), his/her device first publishes this query to a third trusted proxy.
The proxy synthesizes impostor traces through the generator, and generates fake records by extracting ones at the $i$th time interval of impostor traces. The TTP transfers the set of blended data to the LBS server. Due to the indistinguishability, the server responds to all the queries in the set.
The proxy filters the target response from all returned values, and send it back to the user.
At last, the generated fake records are stored in the TTP.
}\label{structure}
\vspace{-0.05in}
\end{figure}

\begin{figure*}
\centering
\includegraphics[width=0.9\textwidth]{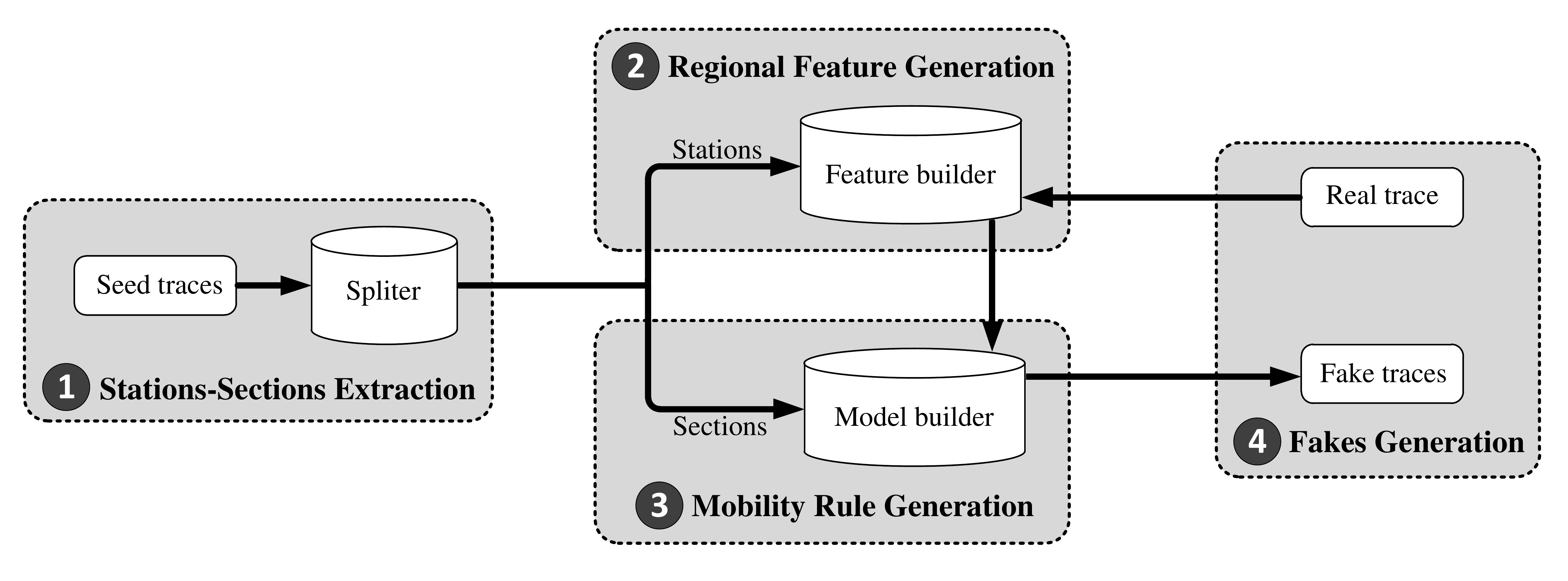}
\caption{
Our scheme includes two modules:
\emph{offline} generator module of synthetic rules (corresponds to the Steps 1-3) and \emph{online} generator module of impostor traces (corresponds to the Step 4).
}\label{scheme}
\vspace{-0.1in}
\end{figure*}

In this section, we present an overview of our scheme to generate dummy traces.
Our approach is deployed in the third trusted proxy (TTP),
which is well accepted in the research community.
The TTP is an architecture with distributed systems so that large-scale data can be stored in it.
We describe the framework and procedures of our method by using the illustration in
Fig. \ref{structure} and its elaborated caption.
The fake records generated before and real trace are stored in the TTP, as done in \cite{Dewri2010Query,Shokri2014Hiding}.
The detailed reason for storing generated fake records will be provided in Section \ref{subsection_fake}.

We discretize time and space by dividing a day into time intervals and dividing the map into fixed-sized grids, obtaining benefits in two aspects:

(1) Keeping the stability of location privacy and utility of service when using LBS.

(2) Partitioning the continuous geographical and temporal data into discrete ones, so that facilitating the mathematical deduction in the following analysis.

As depicted in Fig. \ref{scheme}, there are $4$ steps in our scheme. We first extract the semantic feature of each gird based on the dataset of seed traces.
Each seed consists of stations and sections. We can extract them in the preprocessing phase corresponding to Step 1.
Then by analyzing the temporal distribution of stations of each grid, we compute the similarity between every two grids. We can extract semantic features of these grids by clustering them in Step 2.
In Step 3, we establish the mobility model for users which includes the transition probabilities among grids and the runtime of
passing through grids.
These processes are based on the mobility of the crowd.
Note that the mobility pattern is indeed population-level.
Naturally, the seed traces' privacy can be protected during the processes.
Building the \emph{online} generator module for impostor traces depends on the realization of the former module.
Its procedure refers to Steps 4 in Fig. \ref{scheme}.
In this step, for a real trace, we extract its stations and replace them with records that have similar regional features.
Then we fill in sections among these stations to generate impostor traces.
The attack model in this paper uses a set of training traces to create
a mobility profile for each user in the form of a Markov Chain transition
probability matrix
via
the \emph{knowledge construction mechanism} \cite{synthisiziengsandp2016}.
Having the user mobility profiles and the observed traces, the adversary tries to infer the actual
trace. The estimation is just based on a Bayesian location inference approach \cite{shokri2011quantifying}.

After giving the overview of our scheme, we elaborate two pivotal parts of the
scheme in Section \ref{sec-featurepart1} and Section \ref{sec-tracespart2}, respectively.
Table \ref{notation} presents the list of notations adopted in this paper.

\begin{table}[t]\renewcommand{\arraystretch}{1.5}
\centering
\caption{Table of Notations}\label{notation}\centering
\begin{tabular}{p{1.2cm}|p{6cm}}
\hline Notation & Definition
\\
\hline \hline
\ $r$&a region (location)\\
\hline
\ $t$&a time interval\\
\hline
\ $R$&the set of regions\\
\hline
\ $N_S$&the number of time intervals when generating semantic features\\
\hline
\ $N_M$&the number of time intervals when building mobility model\\
\hline
\ $N_U$&the number of time intervals when sampling locations of user's trace\\
\hline
\ $N_r$(t)&the number of people flow into or flow out of r at t\\
\hline
\ $P_r$&the distribution of people flow into or out of r\\
\hline
\ $C_S$&the semantic class where station $S$ in\\
\hline
\ $G_S$&the semantic similarity graph of regions\\
\hline
\ $G_P$&the transition probability graph of regions\\
\hline
\ $T(t)$&the tensor of the runtime of traveling across regions at $t$\\
\hline
\ $ET_k$&the estimated time of reaching region $k$\\
\hline
\ $d_s(u,v)$&the semantic distance between regions $u$ and $v$\\
\hline
\ $d_g(u,v)$&the geographical distance between regions $u$ and $v$\\
\hline
\end{tabular}
\end{table}

\section{Offline Generator for Synthetic Rules}\label{sec-featurepart1}
In this section, we propose the rules of synthesizing impostor traces.
The processes of generating rules are offline.
We build this generator by the seed traces.
As illustrated in Fig. \ref{scheme}, we divide the module into three parts: extracting stations and sections (Step 1), generating  semantic features of regions (Step 2), and modeling mobility pattern of users (Step 3).

\subsection{Extract Stations and Sections}\label{subsection_station}

We design this module under the framework of spatial-temporal cloaking.
So, we divide the map into fixed-sized grids.
When users' devices upload their trajectories to the trusted proxy, the data are composed of a series of records of GPS coordinates and the corresponding time.
We first standardize the records by matching GPS coordinates to grids, and matching accurate time to time intervals.
For each trajectory, it is a sequence of region-time pairs, denoted by $\langle r,t\rangle$.

The station is not just the simple pause location during the trip, but the phased destination of a user.
There are lots of \emph{stopping} periods of time for vehicles.
These stopping periods can be further divided into two classes:

$\bullet$ Periods of pausing and waiting for moving ahead, e.g., waiting for the traffic light.

$\bullet$  Periods of parking periods, e.g., parking the car and having lunch.

We define the \emph{station} as the record $\langle r,t\rangle$ where and when a vehicle starts or ends the second kind of stopping periods.
Specifically, for a private car driver, the second kind of stopping period is the duration of parking time which exceeds a threshold, and the records of the start and the end of this time duration
are consequently the so-called \emph{stations}.

\begin{figure}[t]
\centering
\includegraphics[width=0.48\textwidth]{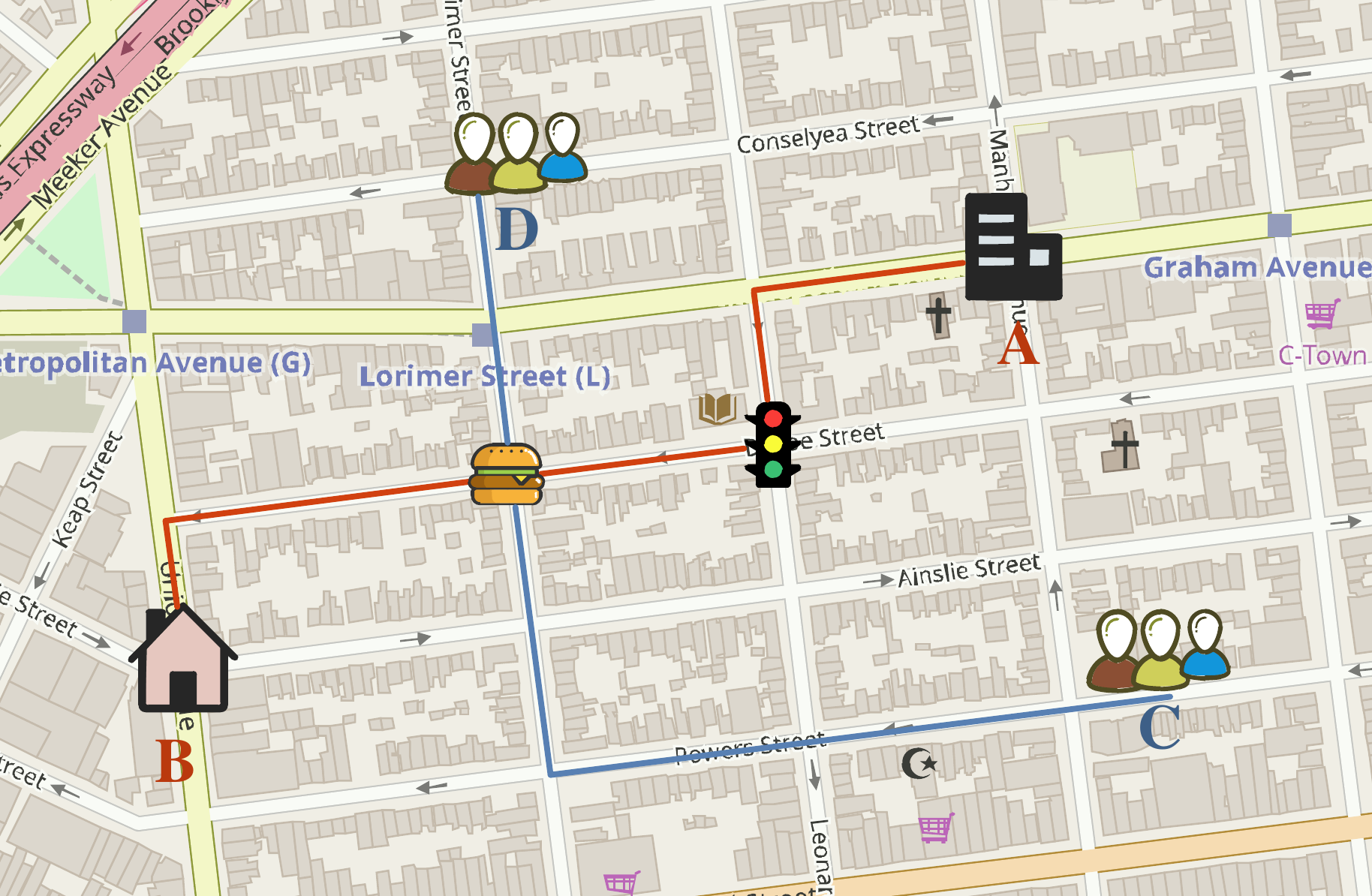}
\vspace{+0.1in}
\caption{
Alice drives back home (location $B$) from the company (location $A$) in the evening, and she stops twice: The first one is the crossroads where she waits for the traffic light;
 the second one is the drive-in restaurant where she gets a burger. The company, drive-in restaurant, and home are stations, and the sections are the traces among them.
Taxi driver Bob picks up passengers at location $C$, and drops them at location $D$.
Here, $C$ and $D$ are stations for him. The corresponding section is the section from $C$ to $D$.
}
\label{DrivingSketch}
\vspace{-0.15in}
\end{figure}

We define the \emph{section} as the targeted trace from a station to the next station. Naturally, a station is the end of the last section, and also the start of the next section. The trace consists of several continuous sections.
Intuitively, stations are the milestones of a trace.
We provide two examples in
Fig. \ref{DrivingSketch}.

\subsection{Generating Semantic Features of Regions}

Our goal here is to generate the semantic features of regions by their visiting patterns.
This process consists of two steps: The first is
to compute the semantic similarities between every two regions; the second is to
cluster regions based on similarities. To this end, we model the distributions of human flow for each region.

In the trajectory of a user, only the stations,
serving as the phased destinations,
make semantic sense,
since the sections between them are usually selected based on
some routing optimization schemes, such as
choosing paths with the minimum costs \cite{TRIP:conf/aaai/LetchnerKH06}.
For example, a man is on a business trip in an unfamiliar city. He intends to go to the company from the hotel.  A rational decision for him is usually to choose the path with the limited time. During the travel process, he does not care about the locations he just passed by. The hotel and company are the only two locations with semantic features for him.
So, we generate the semantic features only by stations. Intuitively, if there is a start or end station of a person's trace in a region, this region has a semantic meaning to him;
while,
 if he just passes through the region, it means nothing to him.
 In fact,
 if we take all the records in the whole trace into account, the non-station records will possibly cause inaccuracy.
In our method, we focus on these stations and neglect the processes between them (process-independent).
For each region grid, we do not care about the movement of people inside and the movement of people just passing through it; we only focus on the temporal distribution of stations in this region.

In Step 2, as illustrated  in Fig. \ref{scheme}, we propose a semantic metric to compare the similarities among stations based on the temporal distributions of the crowd.
In this step, we assign a high similarity value to a pair of regions if the
distributions of \emph{flow-in}s and \emph{flow-out}s of them are really similar, regardless of their geographical distance. For example, the  flow-ins of region $R_A$ and $R_B$ are mostly distributed in the morning, and the majority of flow-outs of them are distributed in the evening. Their distributions of human flow are similar. Although $R_A\neq R_B$, we regard $R_A$ and $R_B$ as semantically similar regions. In this example, $R_A$ and $R_B$ might be workplaces.

\begin{figure}[t]
\centering
\includegraphics[width=0.48\textwidth]{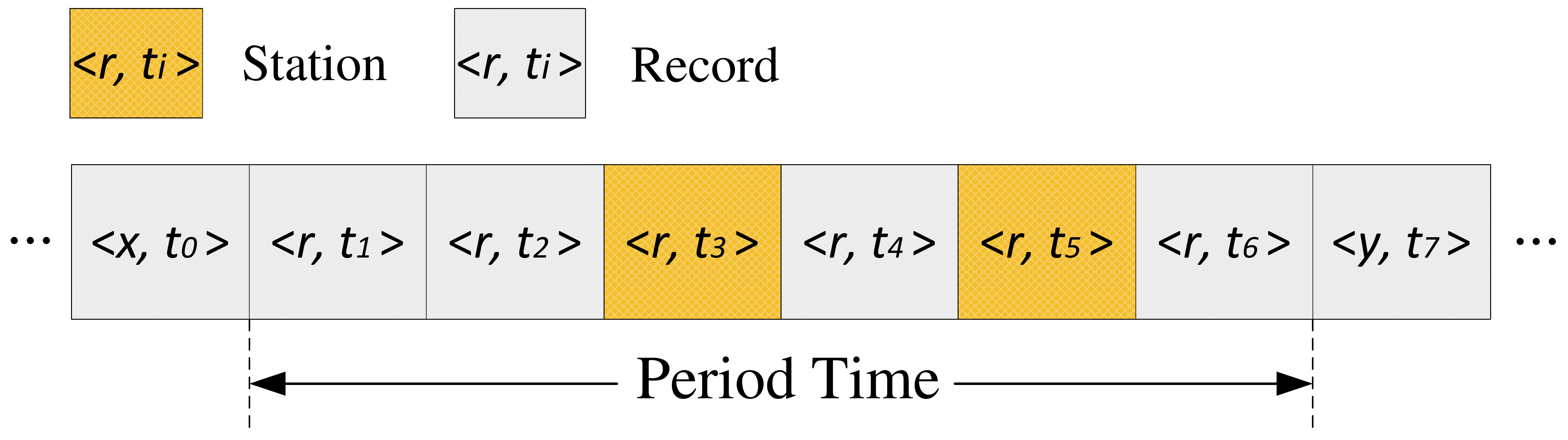}
\caption{An example of computing human flow.}\label{NAdds1}
\vspace{-0.25in}
\end{figure}

The metric of semantic similarity between different regions is to compare the changes of human flow over time.
Thus, we divide a day into time intervals and denote the number of intervals by $N_S$.
In the previous process, we extract stations from seed traces.
Let $N_{in}(r,t)$ and $N_{out}(r,t)$ denote the numbers of human flow-in and flow-out of region $r$ in time interval $t$, respectively. For the different sources of generating human flow, there are two types of methods to calculate $N_{in}(r,t)$ and $N_{out}(r,t)$:

(1) The human flow is based on the statistics of the mobility data of mobile devices' owners (e.g., private car driver).
For example, a man goes to a bar in the evening, which increases $N_{in}$ of this region. In this case, we compute the human flow as follows:
During the period from the time when a client reaches region $r$ to the time he leaves $r$,
if station $\langle r,t\rangle$ in his/her trace is the first station in $r$, we assume that a person enters and visits this region, so $N_{in}(r,t)$ adds $1$. Analogously, if station $\langle r,t'\rangle$ in his/her trace is the last station during this period, he leaves this region, and $N_{out}(r,t')$ adds $1$.
For example, a driver drives into region $r$ from region $x$ at time $t_1$, and parks his/her car to work in his/her company at time $t_3$ (station $\langle r,t_3\rangle$). At the time $t_5$, he leaves his/her company drives home (station $\langle r, t_5 \rangle$), and he reaches the next region $y$ at time $t_7$. So, $N_{in}(r,t_3)$ adds $1$, and $N_{out}(r,t_5)$ adds $1$. The sketch is depicted in Fig. \ref{NAdds1}.

(2) The human flow is irrelevant to the statistics of the mobility data of mobile devices' owners (e.g., taxi driver).
For example, a taxi driver takes a passenger to a hospital. This driver does not contribute to the human flow of this region, but the passenger does. In this case, there must be a signal
that represents the mobility of passengers.
For a record $\langle r,t\rangle$, if the signal indicates that a passenger gets off in region $r$ at time $t$, then $N_{in}(r,t)$ adds $1$. On the contrary, if a passenger gets on,
then $N_{out}(r,t)$ adds $1$.

For each region on the map, after computing the human flow (flow-in and flow-out), we can straightforwardly normalize them by the following equations:

\begin{equation}
\begin{split}
&P_{in}(r,t)=\frac{N_{in}(r,t)}{\sum_{i\leq{N_S}}{N_{in}(r,i)}}, \\
&P_{out}(r,t)=\frac{N_{out}(r,t)}{\sum_{i\leq{N_S}}{N_{out}(r,i)}}.
\end{split}
\end{equation}

We define the semantic similarity based on the Kullback-Leibler Divergence (KLD).
It is a metric of how one probability distribution diverges from the other probability distribution.
We use $KL(P\|Q)$ to represent the KLD between distributions $P$ and $Q$, and $SKL(P\|Q)$ to represent the symmetry KLD:
\begin{equation}
KL(P\|Q)=\sum_{i\leq{N_S}}{P(i)\cdot{\log\left(\frac{P(i)}{Q(i)}\right)}},
\end{equation}
\begin{equation}
SKL(P\|Q)=\frac{KL(P\|Q)+KL(Q\|P)}{2}.
\end{equation}

There are two distributions for every region: distributions of flow-ins and distributions of flow-outs.
For regions $A$ and $B$, we let $P_{in}A$ ($P_{out}A$ ) and $P_{in}B$ ($P_{out}B$) denote the distributions of flow-in (flow-out) of them, respectively.
The semantic distance between $A$ and $B$ is:
\\
\vspace{-0.2in}
\begin{flushleft}
$d_s(A,B)$=\[\alpha\cdot{SKL(P_{in}A\|P_{in}B)}+\beta\cdot{SKL(P_{out}A\|P_{out}B)}.\]
\end{flushleft}
We can compute the semantic similarity between $A$ and $B$ by:
\\
\[Sim(A,B)=1-\frac{d_s(A,B)}{z},\]
where $\alpha$ and $\beta$ (with $\alpha+\beta=1$) represent the importance indexes of flow-in and flow-out, respectively,  and $z$ denotes a normalization constant:
\[z=\max_{X,Y\in R}(d_s(X,Y)).\]

By computing the similarities between every two regions, we can build a graph $G_S$.  Its vertexes are regions, and the weight of an edge  represents the semantic similarity between two vertexes.
We implement the hierarchical clustering algorithm 
on $G_S$ to group regions into distinguishing classes. The regions fall into the same class are similar in the distributions of human flow,
despite their geographical distance from each other.
In other words, the probabilities that people go into and leave them are close in a specific time interval, so they represent the semantic features of regions. For example, if there are two residential areas, people are more likely to leave these areas in the morning and go into them in the evening.
Thus we consider these regions are semantically equivalent.

The process of generating regional semantic features is depicted in Algorithm \ref{computeSimilarity}.

\begin{algorithm} [t]
\caption{Extracting Semantic Features of Regions}
\label{computeSimilarity}
\begin{algorithmic}[1]
\REQUIRE Stations of seed traces
\ENSURE Clusters of regions
\STATE Initiate a weighted graph $G_S$ with regions $R$ as vertices
\FOR {\emph{each region} $r \in R$}
\STATE Compute distribution $P_{in}r$ and $P_{out}r$
\ENDFOR
\FOR {\emph{each region} $r_1 \in R$}
\FOR {\emph{each region} $r_2 \in R$}
\STATE Compute the semantic distance between $r_1$ and $r_2$, i.e., $d_s(r_1,r_2)$,  by:
\\ \vspace{-0.2in}\[\alpha\cdot{SKL(P_{in}r_1\|P_{in}r_2)}+\beta\cdot{SKL(P_{out}r_1\|P_{out}r_2)}\]
\STATE Compute the edge weight between $r_1$ and $r_2$ by:
\\ \vspace{-0.1in} \[w(r_1,r_2)=1-\frac{d_s(r_1,r_2)}{z}\]  \vspace{-0.1in}
\ENDFOR
\ENDFOR
\STATE Return: Cluster regions by implementing Hierarchical Clustering algorithm on $G_S$
\end{algorithmic}
\end{algorithm}

\subsection{Modeling Mobility Pattern}

We build the model of users' mobility pattern in this subsection. It is a location-dependent
first-order Markov chain \cite{Krumm2008A} on the set of regions.
For mobility patterns of individuals usually vary with time,
we partition a day into several time intervals.
Denote the number of time intervals as $N_M$.
The mobility profile is $\langle G_P(t),T(t)\rangle$ of a given time $t$, which will be elaborated later.

In a time interval $t$, $G_P(t)$ is a weighted directed graph which depicts the transition probability between two regions.
It is established based on the Markov chain, which means the position of a user is just related to the last position in his/her trace.
The vertexes of $G_P(t)$ are the regions on the map, and the weight of edge from $r$ to $r'$ is $-\log{P(r'|r)}$,
where $P(r'|r)$ is the probability that users move to region $r'$ from region $r$ in the time interval $t$. If $r'$ is not adjacent to $r$, $P(r'|r)$  equals $0$.
The reason for setting $-\log{P(r'|r)}$ as the weight is that
the process, finding the trace with the $k$th maximum probability, can be transformed into the process of finding the $k$th shortest path in $G_P(t)$.

For a given time $t$, $T(t)$ is a three-dimensional tensor which represents the runtime of passing through regions. The entry $T(r_B,r_A,r_C)$ is the runtime of traveling across region $r_B$ from $r_A$ to $r_C$ in this time interval.
For example, region $r_A$ and $r_C$ are adjacent to $r_B$. In time $t$, a user leaves region $r_A$ at time $t_A$, and enter $r_B$. Then he reaches region $r_C$ at time $t_C$.
We record the entry as:
\begin{equation}
T(r_B,r_A,r_C)=t_C-t_A.
\end{equation}
However, if region $r_X$ is not adjacent to region $r_B$, it holds that
\[\sum\nolimits_{r\in{R}}T(r_B,r_X,r)=0,\]
where $R$ is the set of all regions.

When users are traveling between two stations, they usually choose the way with the minimum cost (time or fuel consumption).
Between the same starting and destination locations, in the same interval, users usually choose the similar path, and the traveling time is similar too.
We model the mobility pattern in a population-level way, where $G_P(t)$ and $T(t)$ are averaged by taking all users into account.
The advantages can be summarized as follows:

(1) The impostor trace performs like ordinary user's trace, and it does not reveal the character of any individual client.

(2) It can protect the individual private trajectories in data processing.

The detailed process is referred to in Algorithm \ref{computeMobility}.

\begin{algorithm}[t]
\caption{Modeling Users' Mobility in Different Intervals}
\label{computeMobility}
\begin{algorithmic}[1]
\REQUIRE Sections between stations of seed traces
\ENSURE Mobility $\langle G_P(t),T(t)\rangle$ for each time interval $t$
\FOR {\emph{each time interval} $t \leq N_M$}
\STATE Initiate a weighted directed graph $G_P(t)$ with regions $R$ as vertices
\FOR {\emph{each region} $r_1 \in R$}
\FOR {\emph{each region} $r_2 \in R$}
\STATE Compute the edge weight by:
\\ $w(r_1,r_2)= -\log{P(r_1|r_2)}$
\ENDFOR
\ENDFOR
\STATE Initial a three-dimensional tensor $T(t)$
\FOR {\emph{each region} $r_1 \in R$}
\FOR {\emph{each region} $r_2 \in R$}
\FOR {\emph{each region} $r_1 \in R$}
\STATE $T(r_1,r_2,r_3)$ $\leftarrow$ the average time of passing through $r_1$ from $r_2$ to $r_3$
\ENDFOR
\ENDFOR
\ENDFOR
\ENDFOR
\end{algorithmic}
\end{algorithm}

\vspace{-0.05in}
\section{Online Generator for Impostors}\label{sec-tracespart2}

In this section, we present the details of our method for synthesizing impostor traces based on the actual location and offline module.
It is depicted in Algorithm \ref{synthesizeTraces}.
When a user proposes a query about a record $\langle r,t\rangle$ to the LBS server, his/her record is updated to the online generator in TTP. In this scenario, the generator extracts a part of the real trace of this user containing the target record and synthesizes impostor traces according to it.
The query is about a record but not a trace. After generating a collection of impostor traces,
we obtain the fake records as follows:

We first determine the time interval $t$ of the actual record (reported to LBS). Then we select the records in the time interval $t$ of impostor traces as the fake ones.

We synthesize impostor traces by utilizing the stations instead of all of the records in the complete trace. The reasons are as follows:

(1) Stations represent the semantic features of activities.

(2) We utilize stations but not all of the records. It means less restriction of the input data.

(3) We are able to synthesize reasonable traces with low complexity.

For users' traces, we divide a day into $N_U$ intervals. It indicates that the locations in the trace are sampled every ${24}/{N_U}$ hours.

\begin{algorithm}[t]
\caption{Synthesizing Plausible Traces}
\label{synthesizeTraces}
\begin{algorithmic}[1]
\REQUIRE  (1) A real record;
(2) Trajectory around the real record;
(3) Fake records generated before;
(4) The number of impostor traces ($n$).
\ENSURE Impostor traces
\STATE Locate real start station $Start$, and real end station $End$
\FOR {\emph{station} $s_i$ $\in$ \{station $|$ stations in the real trace from $Start$ to $End$\}}
\IF {$s_i$ has fake records generated before}
\STATE The set of candidate fake records $\leftarrow$ \{fake records generated before\}
\ELSE
\STATE The location $l$ of $s_i$ $\in$ semantic class $C_l$
\STATE The set of candidate fake records $\leftarrow$ \{records $\langle r,t\rangle$ $|$ $r$ $\in$ $C_l-l$, $t$ = time of $s_i$\}
\ENDIF
\ENDFOR
\FOR {\emph{every two successive stations} $S$, $E$ $\in$ \{station $|$ stations in the real trace from $Start$ to $End$\}}
\FOR {\emph{each candidate fake location} $S'$ of $S$}
\FOR {\emph{each candidate fake location} $E'$ of $E$}
\STATE $d_g(S',E')$ $\leftarrow$ the geographical distance between central points of $S'$ and $E'$
\STATE Add an edge between $S'$ and $E'$ with weight $w$, \\
$w$ $\leftarrow$ $-|d_g(S,E)-d_g(S',E')|$
\ENDFOR
\ENDFOR
\STATE Match candidate fake records of $S$ and $E$ in the bipartite graph by Kuhn-Munkres algorithm
\STATE Choose $n$ pairs of fake records whose geographical distances are the top $n$ similar to $d_g(S,E)$
\FOR {\emph{every pair fake records} $f_S$ of $S$ ($\langle s_S,t_S\rangle$) and $f_E$ of $E$ ($\langle s_E,t_E\rangle$)}
\STATE Time interval $t$ $\leftarrow$ $\frac{t_S+t_E}{2}$  
\STATE Impostor trace $\leftarrow$ $k$th shortest path from $f_S$ to $f_E$ in the graph $G_P(t)$
\FOR {the \emph{$i$th record} in the impostor trace}
\STATE Compute estimate time $ET_i$ of reaching region $i$
\STATE Compute the cloaking time of $ET_i$
\ENDFOR
\ENDFOR
\ENDFOR
\end{algorithmic}
\end{algorithm}

\subsection{Transforming Stations into Fake Ones}\label{subsection_fake}

In this subsection, we generate fake stations based on the target records and side information.
As presented in Section \ref{sec-overview}, the real trace and the fake records generated before are stored in the TTP.
A user proposes a query about the record in the $i$th time interval to the TTP.
The TTP first extracts two parts of information from stored data:
The first part is the user's trajectory during the period from $15$ hours before the time of the target record to $15$ hours after it.
We will explain why we choose $15$ hours as the period in the next paragraph.
The second part contains the fake records generated before.
We divide stations into two categories, the \emph{special} ones and \emph{ordinary} ones. For a station, if we have generated fake records for it in the previous procedures of synthesizing impostors, we regard it as a special station. Otherwise, if it does not have corresponding fake ones, it is an ordinary station.

\begin{figure}
\centering
\includegraphics[width=0.4\textwidth]{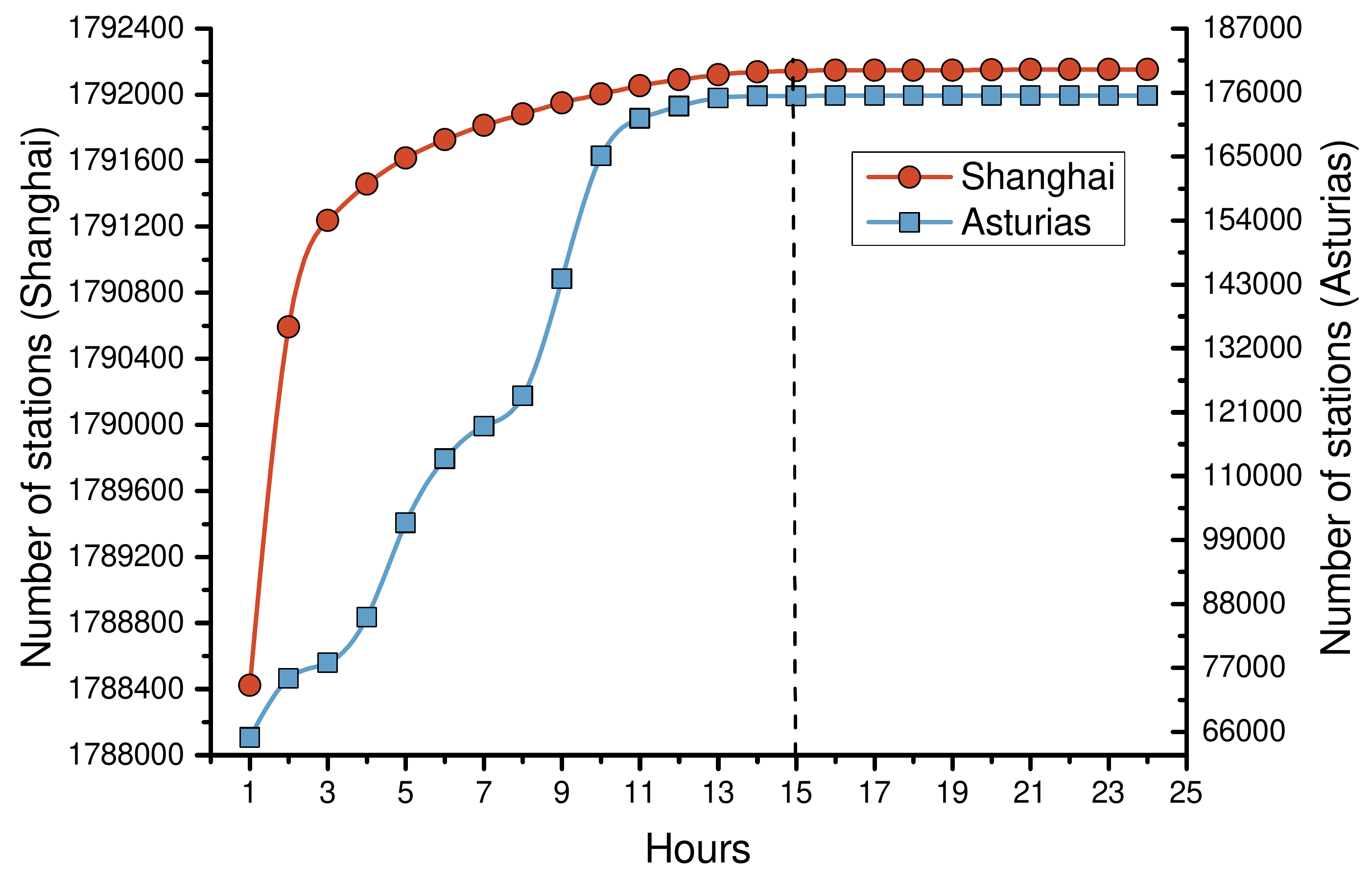}
\caption{
The cumulative distribution function of the number of stations with time.
}
\label{15hours}
\vspace{-0.1in}
\end{figure}

The process of transforming stations into fake ones begins with extracting a part of the real trace.
This part of the trace is user's real trajectory from the start station to end station.
It acts as the template of impostor traces which will be synthesized in our framework.
There are two rules for selecting the start and end stations under different conditions:
(1) In the trajectory from the stored data,
if there are special stations before (after) the target record, we select the special one which is nearest to the target record in the time dimension as the start (end) station. (2) If there is no special station before (after) the target record, we select the ordinary one which is nearest to the target record in the time dimension as the start (end) station.
It is worth mentioning that if there is no station after the target record, we regard the target one as a station.
So, we should find stations in the period around the time of the target record as much as possible. We randomly choose traces of $2000$ vehicles in $30$ days from the dataset of Shanghai, and traces of $50$ vehicles in $100$ days from the dataset of Asturias. Then, we count the time intervals between every two adjacent stations. The cumulative distributions functions of the number of stations are elaborated in Fig. \ref{15hours}. It is evident that the period of $15$ hours can cover most adjacent stations.
This is the reason why we choose $15$ hours as the time period.

\begin{figure}[t]
\centering
\subfigure[$4$ combinations of start and end stations:
(1) Special start station, and special end station.
(2) Special start station, but ordinary end station.
(3) Ordinary start station, but special end station.
(4) Ordinary start station, and ordinary end station.
]
{
\includegraphics[width=0.48\textwidth]{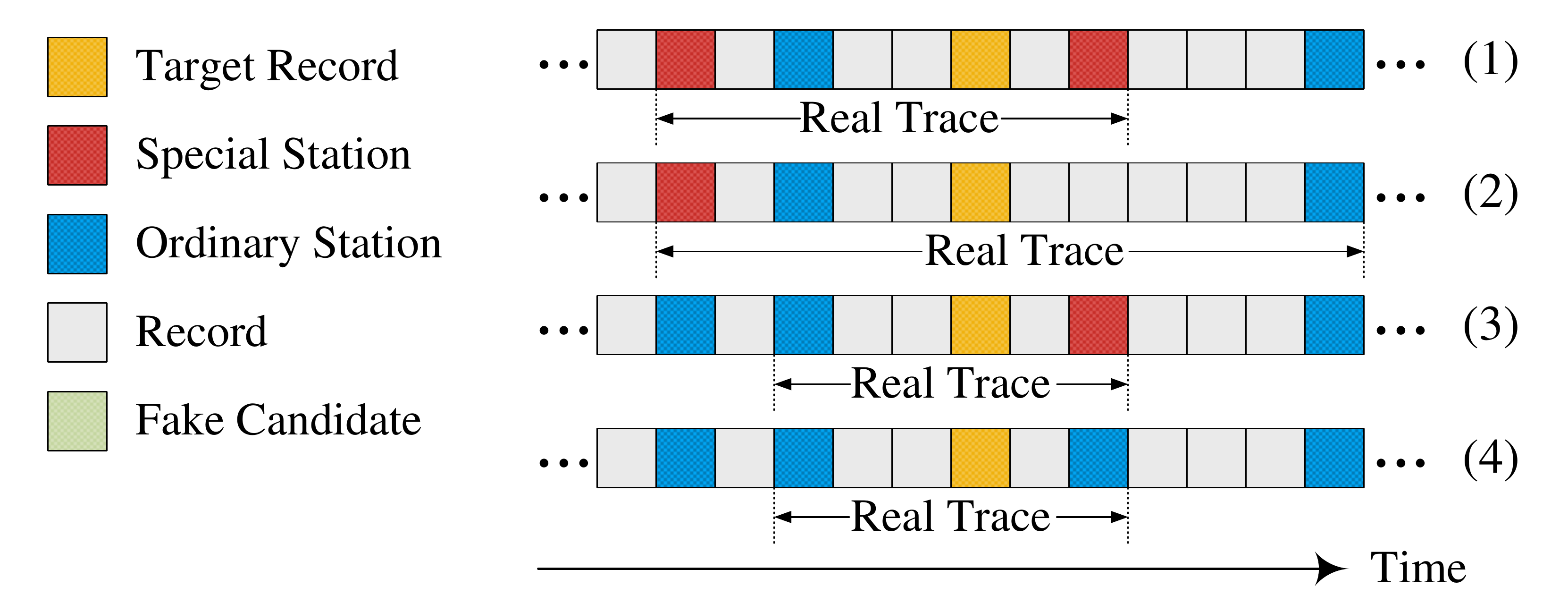}
}\label{fakeStartEnd}
\subfigure[An example of transforming stations into fake ones]
{
\includegraphics[width=0.48\textwidth]{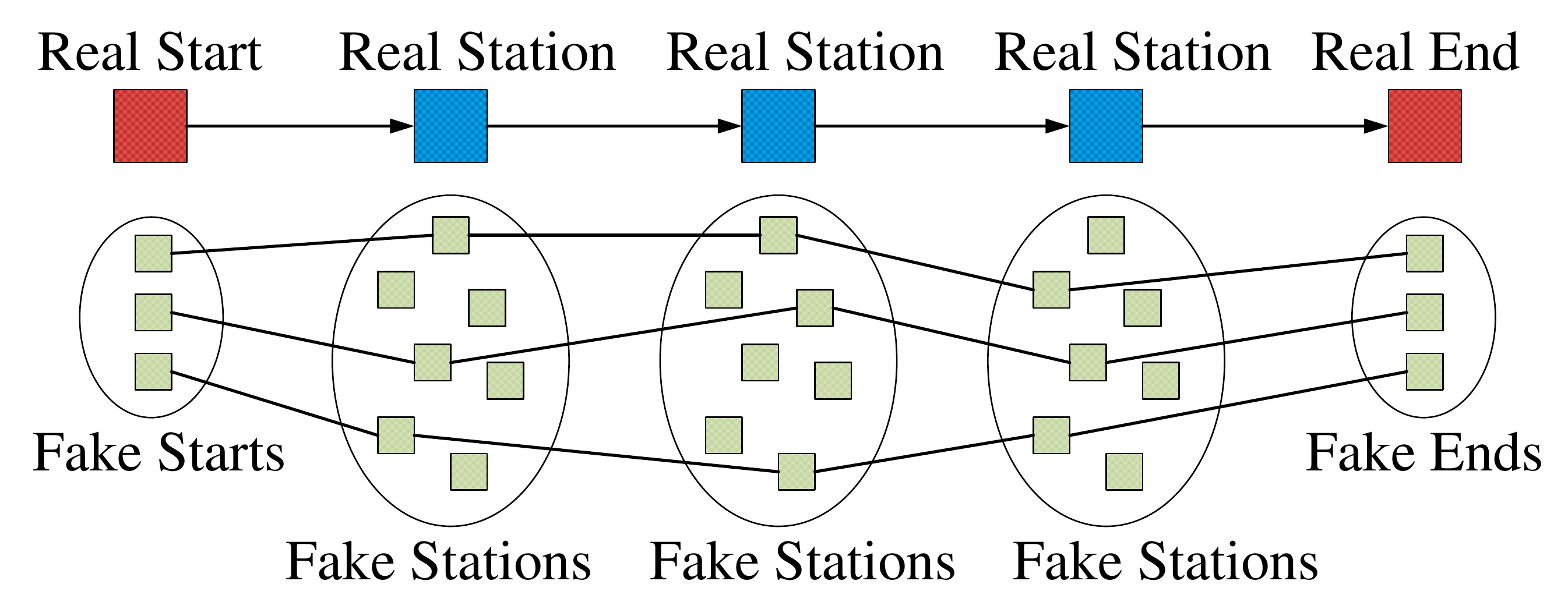}
}\label{bigraph}
\caption{The sketch of transforming stations into fake ones.}
\label{transform2fake}
\vspace{-0.1in}
\end{figure}

After receiving the target record, the TTP first identifies stations in the trajectory (from stored data) according to the definition of stations in Section \ref{subsection_station}.
Then the TTP selects the start and end stations to extract the part of the real trace.
As depicted in Fig. \ref{transform2fake}(a), due to the different states of stations, we have $4$ combinations of the start and end stations.

For a special station, in this process, the fake candidate records are the same with ones generated before.
For an ordinary station, to keep the semantic similarity between the fake and actual stations, the fake station should be extracted from the same semantic class as the actual one.
For any station $S$ in a trace, its location $X$ corresponds to the regional semantic class $C_X$.
In order to prevent the leakage of the original information, we remove $X$ from $C_X$ before picking the fake station. So, we choose the fake location $X'\in C_X-X$. In this case, the fake candidate stations are the records

\vspace{-0.1in}

\[\left\{\langle r, t\rangle | r\in C_X-X, ~t= \mbox{time of  }S \right\}.\]

\vspace{-0.05in}

It is noted that there may be some other stations in the trace from the start to end.
We synthesize fake the sub-traces between every two continuous stations.
For any station in this trace,
it has the corresponding fake candidates, and its fake stations are selected from them.
The selection of fake stations is not randomized.
It utilizes the rule that the distance between two continuous candidate fake stations should correspond to the travel time of real trace.
Otherwise, the fake stations can be identified easily by attackers. For example, if two fake stations are $50\mbox{km}$ apart, but the runtime between their corresponding real stations is just $10$ minutes, it is manifestly unrealistic.
Intuitively, we intend to ensure that fake stations are semantically and geographically similar to the actual pairs.

To keep the utility of service, we partition the map into small grids, e.g., of area $1\mbox{km}\times1\mbox{km}$.
We define the distance between regions as the distance (in a straight line) between two central points of them.
For every two continuous stations in the real trace, the first one is the start station $S$, and the second is the end station $E$.
We compute the distance between every two regions from two sets.
One is the locations set $F_S$ of fake candidates of $S$, and the other is the locations set $F_E$ of fake candidates of $E$.
We can build a bipartite graph whose vertices are divided into two disjoint and independent sets $F_S$ and $F_E$,
respectively.
The weight of the edge connects $S'\in F_S$ and $E'\in F_E$ is calculated by:
\[w=-|d_g(S,E)-d_g(S',E')|,\]
where $d_g$ represents the geographical distance.
After building the graph, we utilize the Kuhn-Munkres algorithm
to match the region from $F_S$ to the one from $F_E$, and keep the maximum sum of weights \cite{Zhu2012Group}.
To generate $n$ impostor traces, we choose $n$ pairs of regions, whose geographical distances are the top $n$ similar to the real.
Remark that the impostor trace should have the same number of stations with the real trace.
For example, if $n=3$, the sketch of transforming stations into fake ones is shown in Fig. \ref{transform2fake}(b).
After generating fake records for an end station, its candidate set turns to be
fake records generated just now, and it will be applied to generate fake records for the next station.
This approach can keep the geographical similarity between imposters and real traces.

\subsection{Complementing Impostor Trace}

A trace can be represented by a sequence of records. In this subsection, we fill in the trace among fake stations.
For a section from the start station to end station,
any random walk appears to be logical. However, if a vehicle is in the wrong direction of a road, it can easily be debunked. We fill in the section based on the mobility model $G_P(t)$ in a specific time interval $t$ ($t\leq{N_M}$).
For a pair of the start location $S$ and end location $E$, our goal is to find out the trace ($S,r_1,r_2,r_3,...,r_n,E$) with the largest probability. The process of this step can be formulated by:
\begin{eqnarray*}
\arg\max_{trace}P(trace|S,E)&=&\arg\max_{trace}(P(S)\cdot{P(r_1|S)}\\
&&\times{\prod_{i=1}^{n-1}P(r_{i+1}|r_i)}\cdot{P(E|r_n)})
\end{eqnarray*}
For the value of $P(S)$ is a constant for a specific start location, the formula above can be shortened as:
\begin{equation}
\arg\max_{trace}({P(r_1|S)}\cdot{\prod_{i=1}^{n-1}P(r_{i+1}|r_i)}\cdot{P(E|r_n)})
\end{equation}
However, consider a situation, there is a strong attacker who can construct the knowledge of user's mobility in advance. If the impostor traces generated by our model are always based on the maximum transition probability, the attacker can filter out the impostor traces easily, because the probability of real traces is not always the maximum value.
Hence, we choose the trace with the $k$th greatest probability as the fake.
To increase the robustness of our algorithm, we add some randomness to the selection of the value of $k$. The probability of $k$ is dependent on the statistic of the real data.

In our weighted directed graph of regions, given the start and end regions, we can select the trace with the
$k$th greatest probability by computing the $k$th shortest trace between two regions in the graph $G_P$.
We find the $k$th shortest path based on the Dijkstra's algorithm and $A^*$ algorithm \cite{Eppstein1994Finding}.
The weight of the edge is $-\log(P(r'|r))$, so that the distance in the graph of a trace ({$S,r_1,r_2,r_3,...,r_n,E$}) is
\begin{center}
$
-\log\left(P\left(\frac{r_1}{S}\right)\cdot{\prod\nolimits_{i=1}^{n-1}P\left(\frac{r_{i+1}}{r_i}\right)}\cdot{P\left(\frac{E}{r_n}\right)}\right).
$
\end{center}
The $k$th shortest trace is the trace with the $k$th greatest probability.

\subsection{Add Timestamp}

\begin{figure}[t]
\centering
\includegraphics[width=0.45\textwidth]{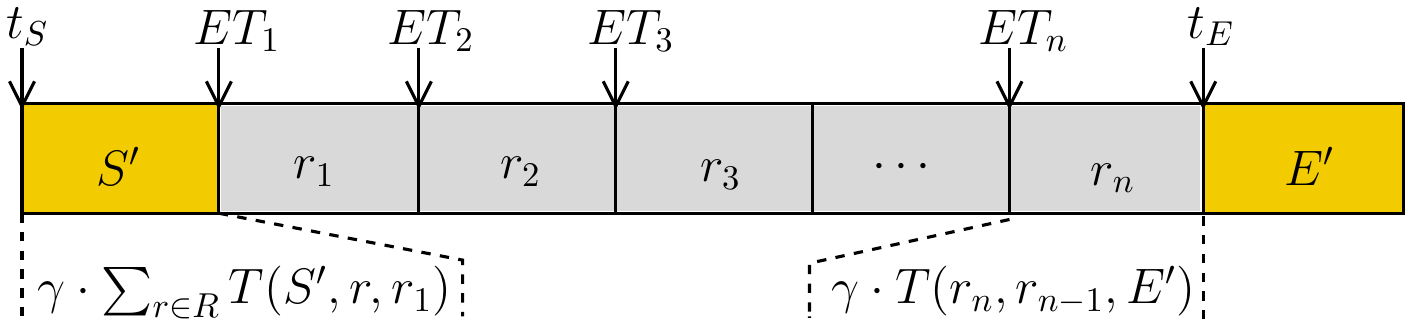}
\caption{An illustration of adding timestamp.}
\vspace{-0.1in}
\label{addTimestamp}
\end{figure}

After generating the impostor trace ($S',r_1,r_2,r_3,...,r_n,E'$), we add timestamps to each region in the trace.
As illustrated in Fig. \ref{addTimestamp}, the start time is $t_S$ and end time is $t_E$ according to the real trace.
We assign the time interval $t$ as the interval including the time $(t_S+t_E)/2$.
The runtime of traveling through every region along the impostor trace is the statistical time multiplied by $\gamma$.
We compute $\gamma$ and the estimated time $ET_k$ to arrive at region $k$ by the following procedures:
\vspace{-0.1in}
\begin{eqnarray*}
   \Phi&=& \sum_{r_i\in{r}}T(S',r_i,r_1)+\sum_{i=1}^{n-2}T(r_{i+1},r_i,r_{i+2})\\
 &&+T(r_n,r_{n-1},E'),\\
 \Psi&=&\sum_{r_i\in{r}}T(S',r_i,r_1)+\sum_{i=1}^{k-1}T(r_{i+1},r_i,r_{i+2}),\\
 \gamma &=&  \frac{t_E-t_S}{\Phi}, \\
 ET_k&=& t_S+\gamma\cdot \Psi.
\end{eqnarray*}
The time we estimate is accurate to second order. Therefore, we need to transform it into the interval as LBS providers' requirement.
We can compute the cloaking time interval of region $k$ in the impostor trace as $\left\lceil\frac{ET_k}{N_U}\right\rceil$.

\section{Evaluation}\label{sec-evalu}

\subsection{Experiment Setup}

In this section, we implement our method on the real traces of two datasets. One is the dataset of taxis' traces in Shanghai of China \cite{shanghai},
and the other is the dataset of private cars' traces in Asturias, Spain \cite{dartmouth-campus-20160808}.
The vehicular trajectories data in Shanghai were collected from $13693$ taxis from April 1st, 2015 to April 30th, 2015. They were recorded every few seconds in $30$ days. The raw dataset of Shanghai is a series of records, which contains the latitude, longitude, time and the signal of whether carrying passengers.
The original data in Asturias is the GPS traces for one year collected from $142$ cars,
and records are reported with an interval of $30$ seconds. The Asturias' dataset is a series of records containing the latitude, longitude, and time.
The experiments are implemented based on the different scales of maps in Shanghai and Asturias, respectively.

\begin{figure}
\centering
\includegraphics[width=0.4\textwidth]{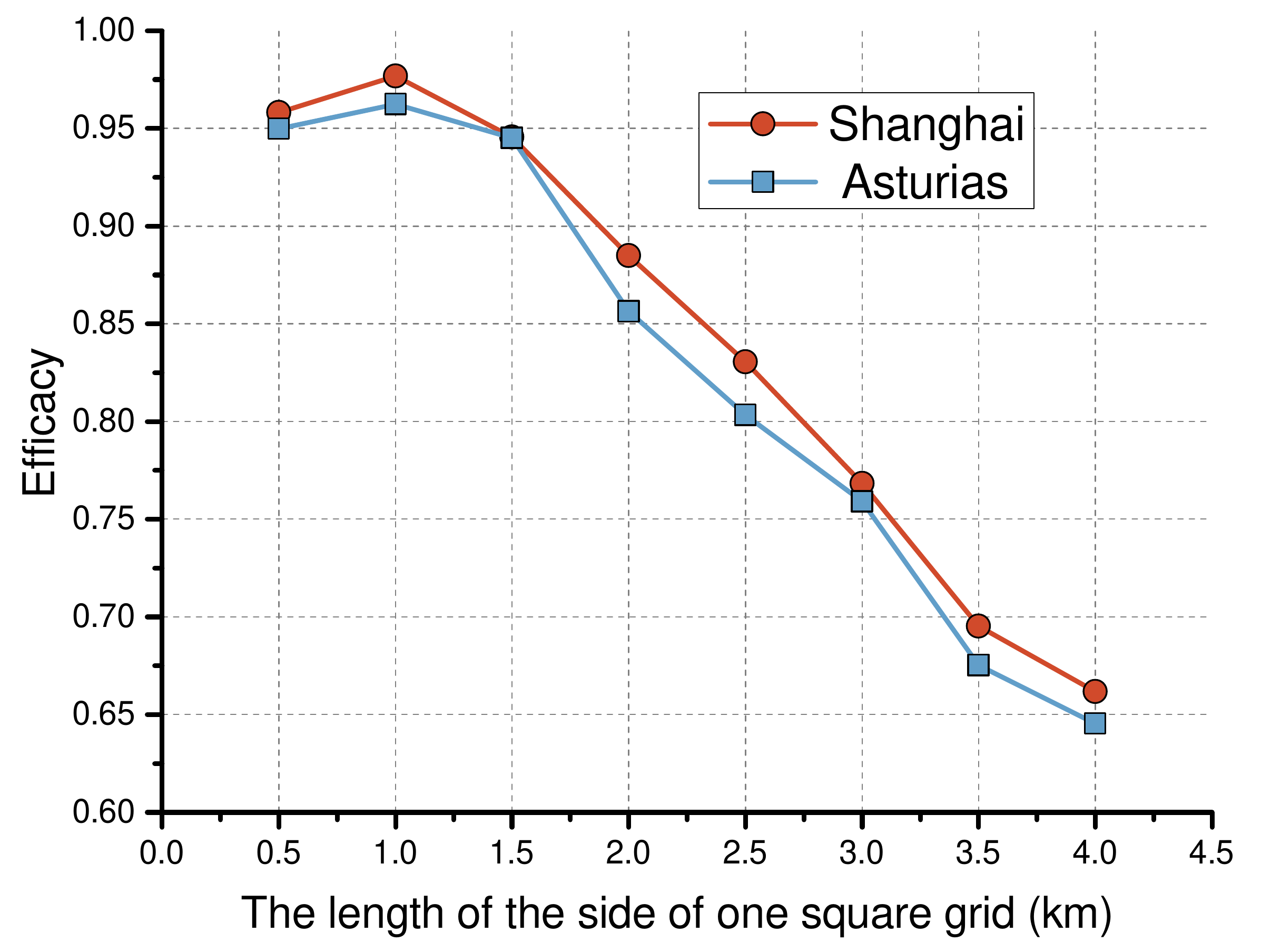}
\caption{
The relationships between the efficacy and the length of the side of one square grid.
}
\label{gridsize}
\vspace{-0.25in}
\end{figure}

We first preprocess the datasets: Dividing the map into regular size grids, and dividing a day into several time intervals ($N_M$).
To obtain the most proper size of the grid, we record the relations between the different sizes of the grid and the efficacy while synthesizing $10$ impostors for each trace. The result is depicted in Fig. \ref{gridsize}. In these two datasets, when we divide the map into regular $1\mbox{km}\times1\mbox{km}$ grids, we can achieve the best efficacy.
We make an explanation as follows: If we choose a small size to generate a grid,
it might contain
not enough stations for extracting the regional similarity. If we choose a large size, the number of grids in the same cluster is possibly limited.

\begin{figure}[t]
\centering
\subfigure[Dataset in Shanghai]
{
\includegraphics[width=0.4\textwidth]{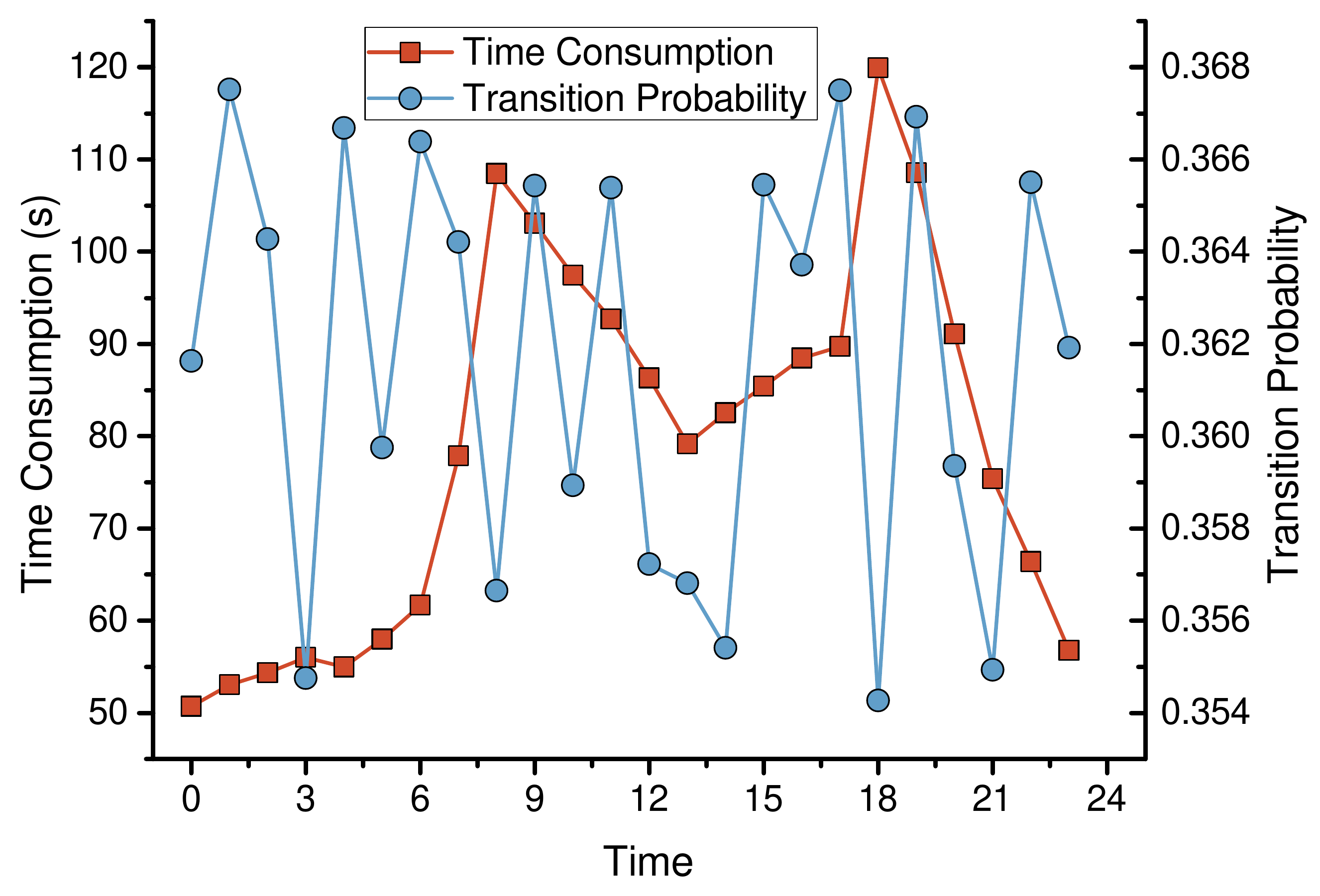}
}
\subfigure[Dataset in Asturias]
{
\includegraphics[width=0.4\textwidth]{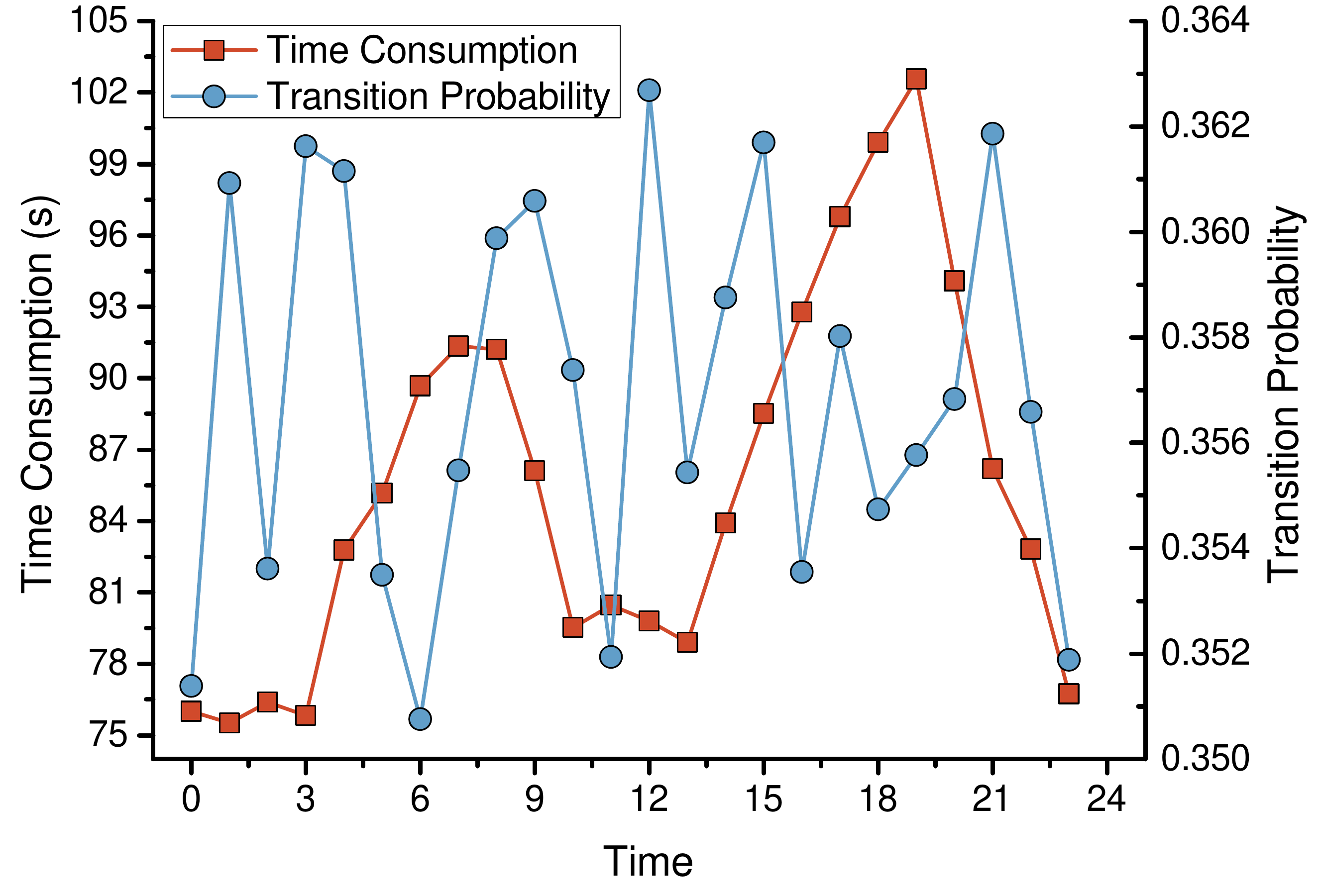}
}
\caption{The change of time consumption and transition probability with time growing.}
\label{c_p_time}
\end{figure}

To choose the appropriate $N_M$, we make statistics of the time consumptions when passing through a grid, and the transition probabilities between grids in the different time for two datasets.
In every hour, we randomly choose $100$ grids and calculate the average time consumption when passing through them. We also randomly choose $100$ pairs of grids to calculate the average transition probability between them.
As depicted in Fig. \ref{c_p_time}, the result shows that the transition probability of one dataset fluctuates around a constant value. However, the time consumption of traveling through a grid varies from hour to hour.
To keep the accuracy, we build the mobility model every hour in a day. This means that $N_M=24$.

For the dataset of taxis' traces in Shanghai, if a taxi stops for a while and picks up passengers (lets passengers off), stations are the start and the end records of these time periods. It is noteworthy that if the time and regions of them are the same, they will be regarded as the same station.
The dataset in Shanghai has the signal of whether carrying passengers. If the signal of a record changes from $0$ to $1$, it represents a passenger getting on at this time. So there is a passenger leaves this location.
On the contrary, when a passenger is getting off, he goes into this location. In the experiment, we extract $11569124$ stations from the dataset.

\begin{figure}
\centering
\includegraphics[width=0.4\textwidth]{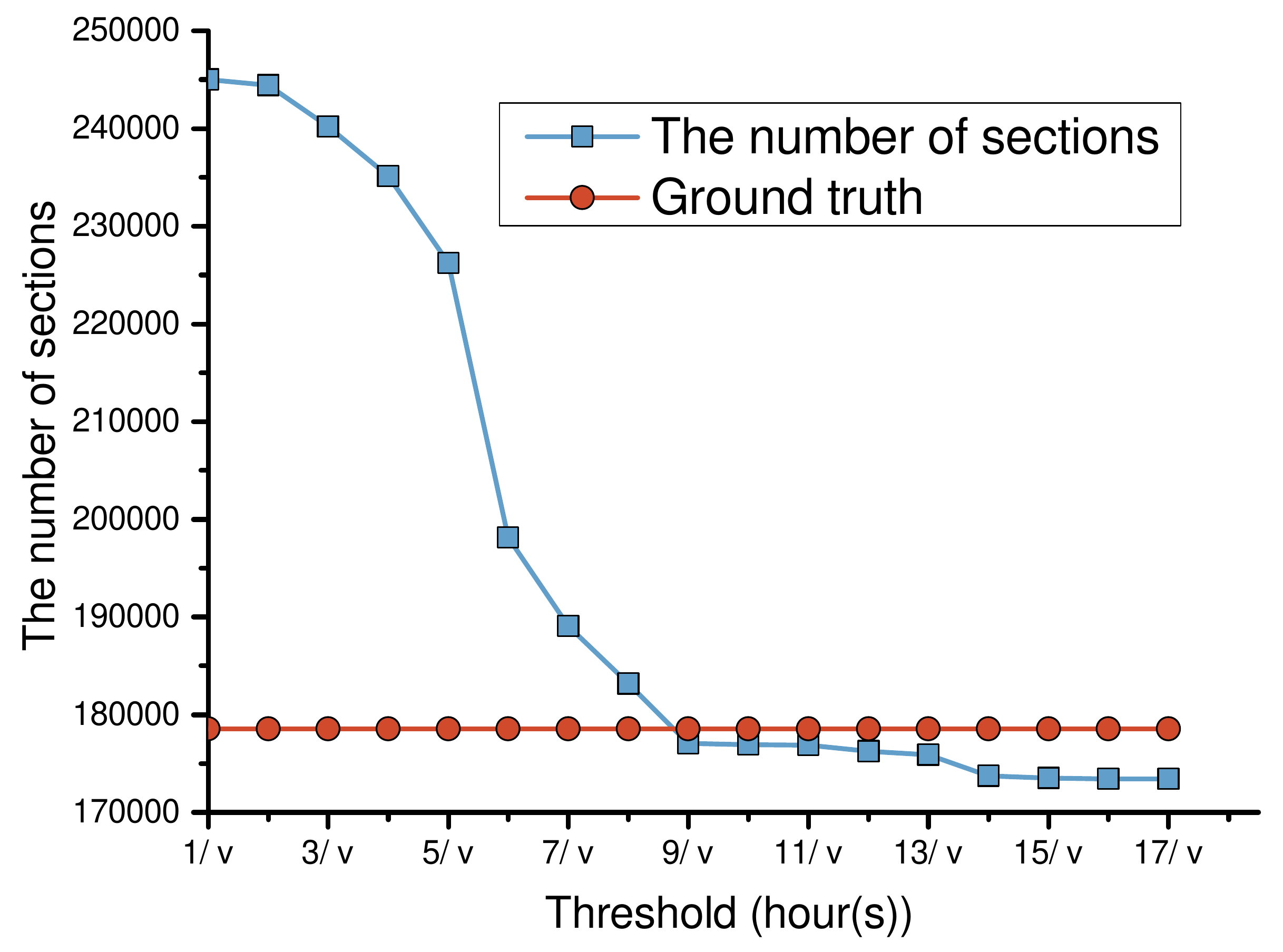}
\caption{The relations between the ground truth and the number of stations under the different thresholds. }
\label{station_threshold}
\end{figure}

For the dataset of private cars in Asturias, we assume that the start and end of a period are stations, if the period of a vehicle parking in a grid is longer than a threshold.
The dataset in Asturias are recorded every $30$ seconds (accurate time, longitude and latitude), and we can judge whether a car stops from it. The time of a car waiting for the traffic light is almost less than $300$ seconds. Accordingly,
 if a car stops more than $300$ seconds, the start and the end records of this stopping period are stations. We extract $178537$ stations from $50$ vehicles in $100$ days.
We regard the number of these stations, i.e. $178537$,
as the ground truth in Fig. \ref{station_threshold}.
However, our method is implemented in the TTP. The third trust proxy just stores traces data of time and grids, and the TTP cannot judge whether a car stops in a grid. So we extract stations by a threshold in our method.
Note that the thresholds are variable in different grids and time intervals.
We denote the average speed of region $r$ in time interval $t$ as $v(r,t)$ km/hours.
When using different thresholds, we make statistics of the numbers of stations from $50$ vehicles in $100$ days. The result is depicted in Fig. \ref{station_threshold}.
So we set the threshold in this region equals $\frac{9}{v(r,t)}$ hour(s).
The first and the last records in this grid are stations, which means that a user goes into and leaves this location, respectively.
Accordingly, we extract $1850979$ stations from this dataset.

\begin{figure}
\centering
\includegraphics[width=0.39\textwidth]{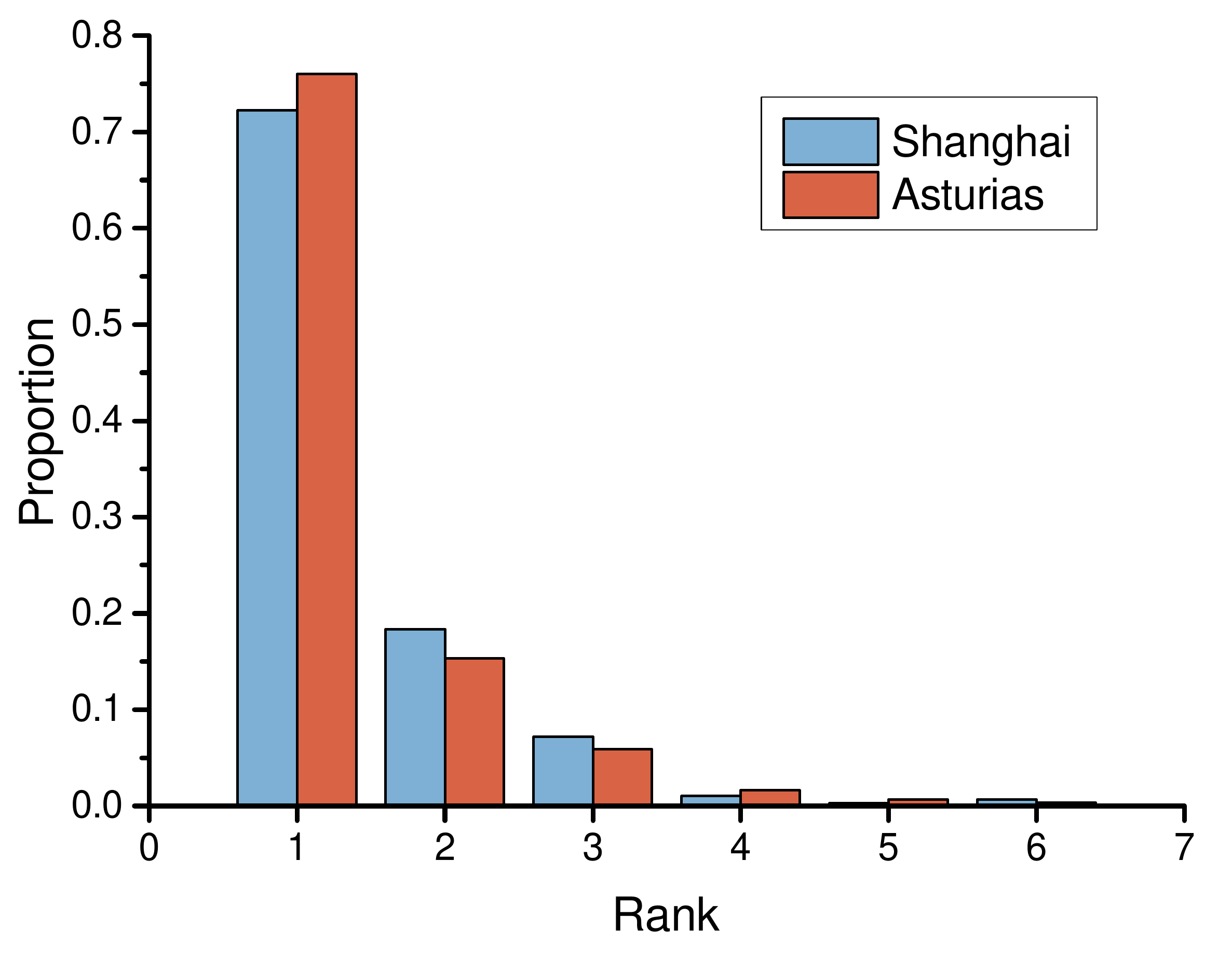}
\caption{The relation between rank ($k$) and proportion.}
\label{traceProbRank}
\end{figure}

In the step of filling the path based on $G_P(t)$, we choose the $k$th shortest path. The probability of $k$ depends on the statistic of the real data. We count the transition probability rank of traces between stations. The relations between the rank ($k$) and proportion of two datasets are both shown in Fig. \ref{traceProbRank}.

The experiments are implemented in Python 2.7.
The configuration of the used computer includes: A CPU Intel(R) Xeon(R) E$5$-$2665$ at $2.4$ GHz, $32$ GB of RAM, $1$ TB of disk space and Windows Server $2008$ R$2$ Enterprise.

\subsection{Efficacy}

\begin{figure}[t]
\centering
\subfigure[Dataset in Shanghai]
{
\includegraphics[width=0.4\textwidth]{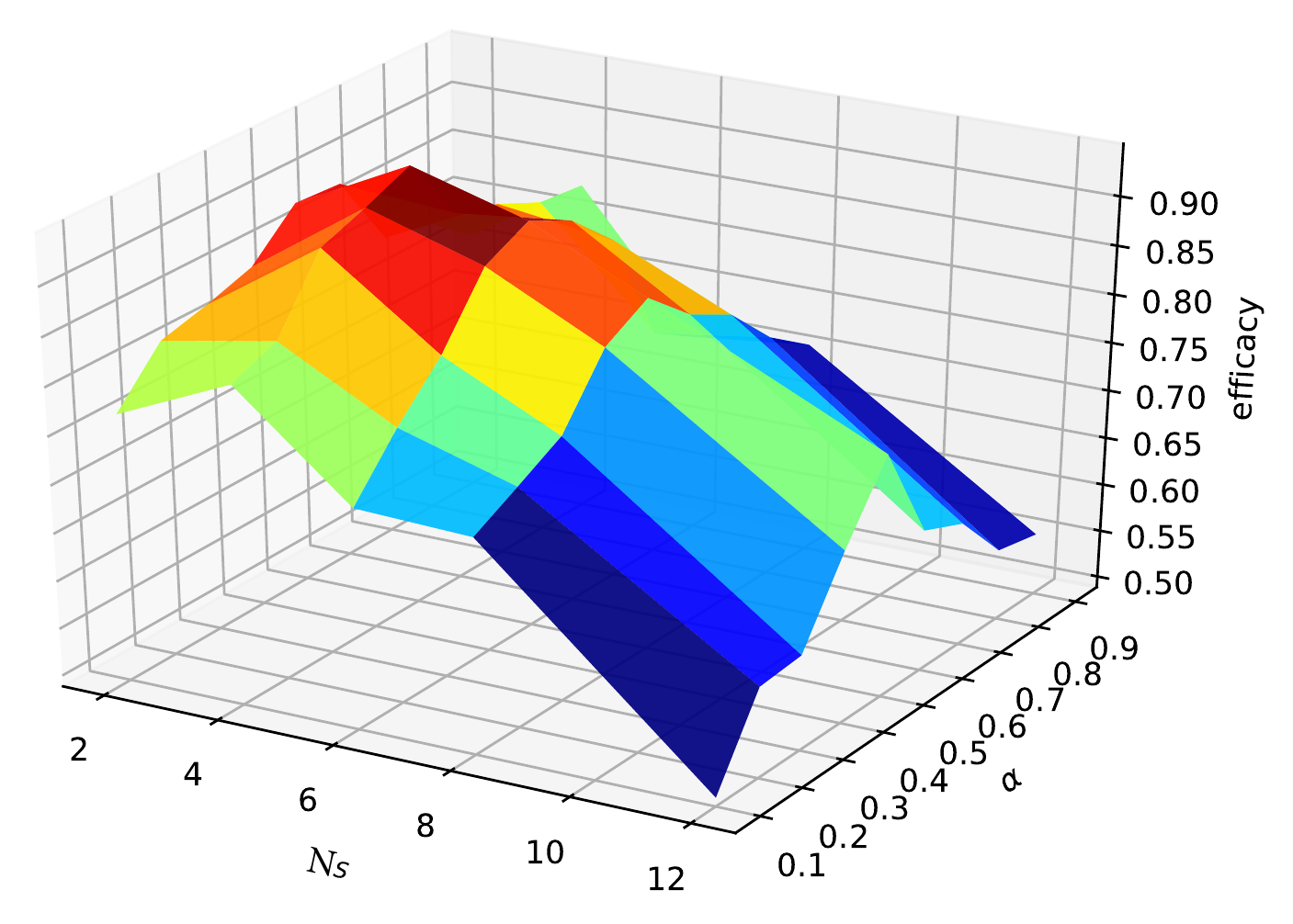}
}
\subfigure[Dataset in Asturias]
{
\includegraphics[width=0.4\textwidth]{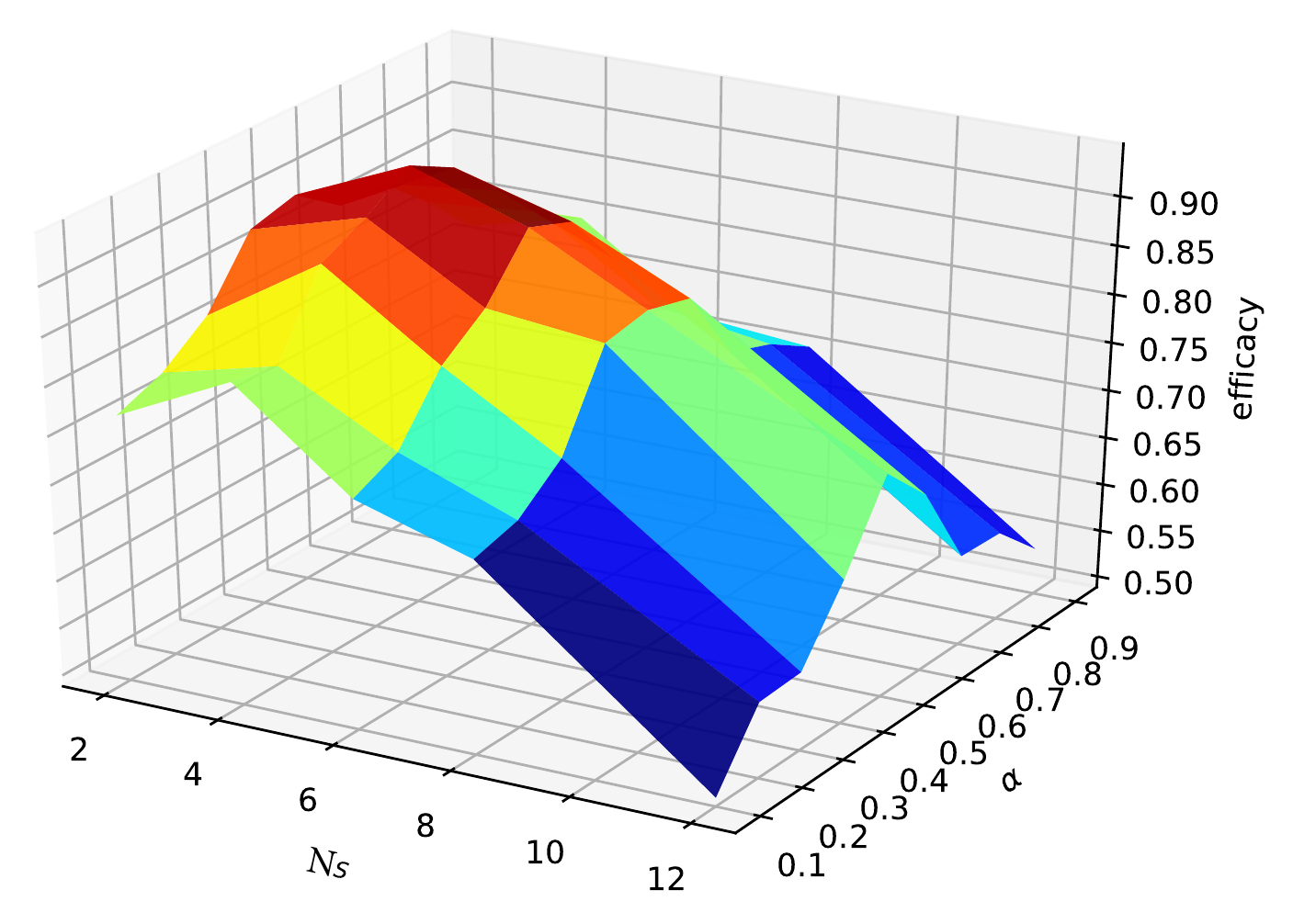}
}
\caption{The Relations of Efficacy, $N_S$, and $\alpha$.}
\vspace{-0.15in}
\label{Pes_alpha_efficacy}
\end{figure}

In this subsection, we measure how effectively our method works by a general inference attack \cite{shokri2011quantifying}.
Specifically, given the user's observed traces and fundamental knowledge needed, the state-of-the-art adversary's attack model filters out the real traces by inference attack.
In this attack model, the attacker uses a set of training traces to create, via the knowledge construction mechanism, a mobility profile for each user in the form of a Markov Chain transition probability matrix. Having the user mobility profiles and the observed traces, the adversary tries to infer the actual traces.
They compute the expected error of an adversary who observes a user's trace and then forms a probabilistic estimation of his/her location. This probabilistic estimation is based on a Bayesian location inference approach.
The preservation efficacy of model is defined as the probability of error to infer the real trace by an attacker.
The higher the probability is, the more effectively the protection mechanism performs.
Consider an exemplary situation. The TTP sends a real trace of Alice and impostor traces together to LBS provider,
and the adversary Bob has access to the LBS queries.
He tries to infer the real trace from blended data by a strong inference attack.
If he finds $2$ real traces from $10$ groups of blended data, the accuracy rate of the adversary is $20\%$, i.e., the preservation efficacy is $80\%$.

In the experiment for testing the efficacy, we adopt the fine-grained traces.
We divide a day into $288$ intervals ($N_U=288$), and input $15000$ seed traces to build the offline generator.
For the dataset in Shanghai and Asturias, the average lengths of seed traces are $3.82$ and $3.58$ intervals, respectively. Maps used in the experiment are the central areas of Shanghai city and Asturias city.
Both of them contain $12\times9$ grids.

To choose the most proper parameters $\alpha$, $\beta$,
 and $N_S$ when generating regional semantic features, we investigate the relations among efficacy, $N_S$, and $\alpha$ in Fig. \ref{Pes_alpha_efficacy} when synthesizing $4$ impostors for each trace. According to the result, we set the importance index $\alpha=\beta=0.5$ and $N_S=4$.

\begin{figure}
\centering
\includegraphics[width=0.4\textwidth]{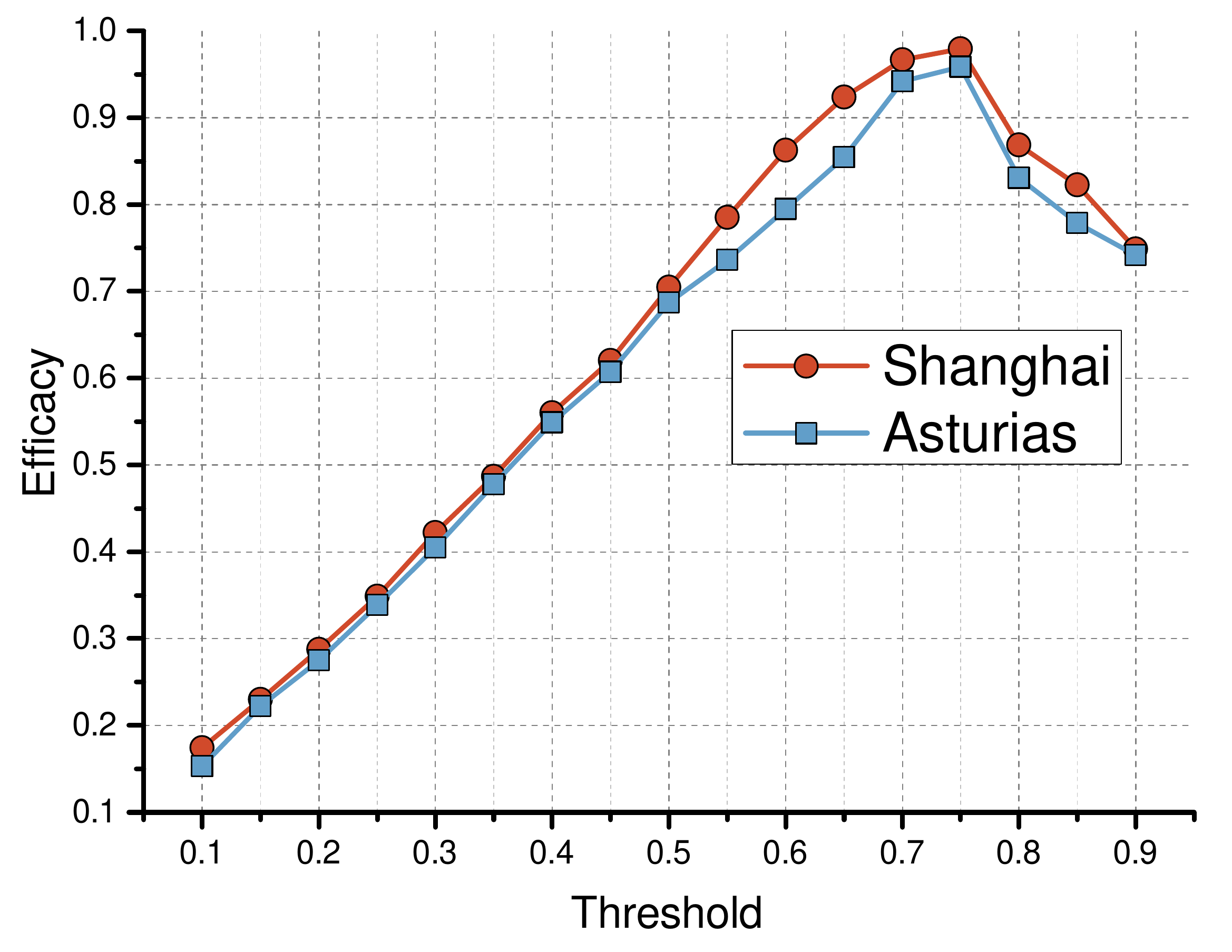}
\caption{The relations between the efficacy and the different thresholds of the hierarchical clustering algorithm.}
\label{threshold_clustering}
\vspace{-0.2in}
\end{figure}

After the process of computing semantic similarity among regions by Algorithm \ref{computeSimilarity}, we set the threshold in the \emph{Hierarchical Clustering} algorithm to be $0.75$, and cluster grids based on the similarity graph.
The relations between the threshold of hierarchical clustering algorithm and the efficacy are shown in Fig. \ref{threshold_clustering}. This gives the reason why we choose $0.75$ as the threshold.
Then we input the test traces to synthesize impostor traces by Algorithm \ref{synthesizeTraces}.

We generate the different numbers of impostor traces ($1$, $4$, $7$, and $10$ per real trace) through $4$ different synthetic generation techniques:

(1) LA-Lppm \cite{suzuki2010user}: Generating dummies around the user in a grid to achieve the restriction of movement consistency and anonymous area. It also makes dummies cross paths with the user to reduce the traceability of the user's location.

(2) RDT-Lppm \cite{krumm2009realistic}: Generating fake locations based on traces. It approaches the realism of actual traces by using probabilistic models of driving behavior abstracted from real traces. It utilizes $3$ characteristics which are derived from a statistical analysis of actual driving traces: realistic starting and ending points, goal-directed routes with randomness, and spatially varying GPS noise.

(3) DAP-Lppm \cite{kato2012dummy}: Anonymizing a user's location based on his/her movements with pauses when using LBSs with mobile devices. It generates dummies that move while stopping at several locations like a real user; the dummies also take into consideration geographical restrictions.

(4) SG-Lppm \cite{synthisiziengsandp2016}: Generating fake traces of the real trace by the semantic features of regions.
It extracts the semantic features through matching every location of every two users. In the process of generating fake traces, it replaces all locations in a user's real trace with ones in the same semantic cluster and generates traces by the Hidden Markov Model.

\begin{figure}[t]
\centering
\subfigure[Dataset in Shanghai]
{\includegraphics[width=0.4\textwidth]{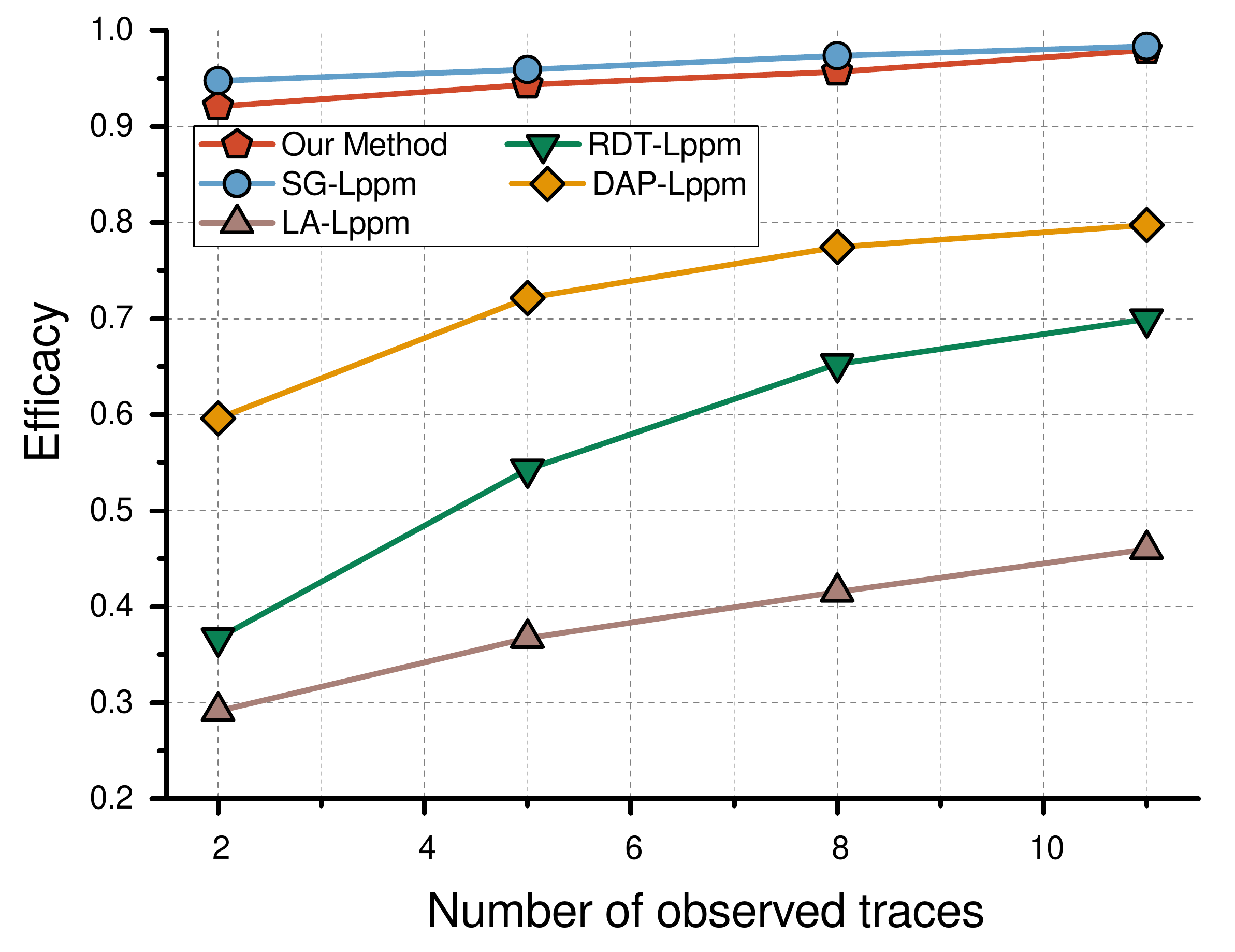}}
\subfigure[Dataset in Asturias]
{\includegraphics[width=0.4\textwidth]{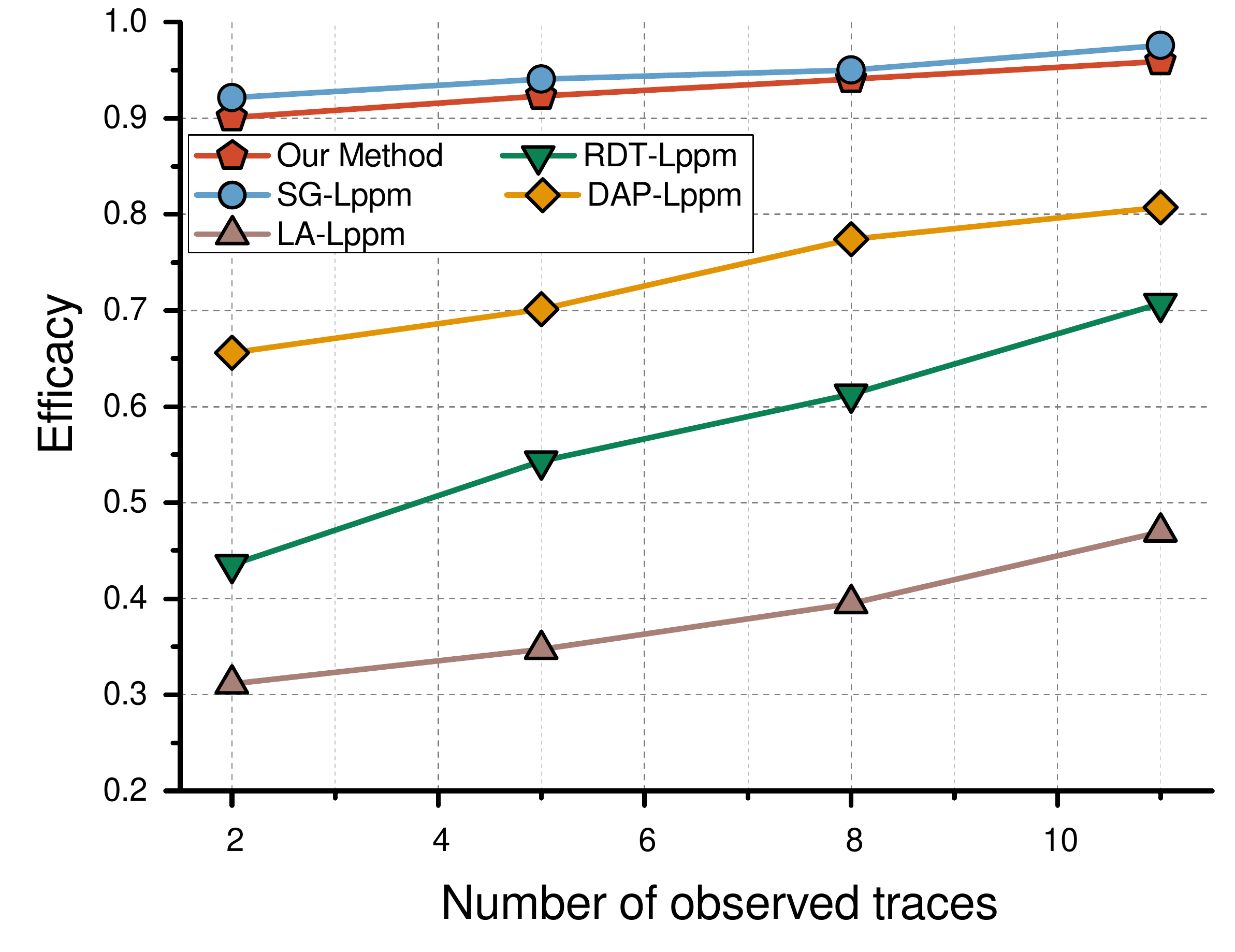}}
\caption{Efficacy of Different Generation Techniques.}
\label{efficacy}
\end{figure}

After that, we input the observed traces which contain the real and impostor traces to the attack model \cite{shokri2011quantifying}, and compute the efficacy of different methods.

Fig. \ref{efficacy} depicts the efficacy of these methods. The result shows that our method and SG-Lppm perform much better than others in these two datasets, achieving the efficacy surpassing $90\%$ when inputting different numbers of observed traces.
When uploading $11$ observed traces for each real trace, our approach can achieve the preservation efficacy of $97.95\%$, and $95.91\%$ in datasets of Shanghai and Asturias, respectively, compared with $98.33\%$ and $97.59\%$ by SG-Lppm.
As a matter of fact, SG-Lppm performs slightly better than ours.
The methods, LA-Lppm, RDT-Lppm, and DAP-Lppm, do not consider the semantic features of traces and have randomness in generating impostor traces.
For the case of the state-of-the-art protection SG-Lppm, it synthesizes traces based on the regional semantics which are extracted from time and space features of human mobility.
They build a straightforward but high-cost module by considering all locations in individual traces.
Compared with SG-Lppm, in our method, a population-level model neglects the latent spatiotemporal correlations among visited locations, then inevitably leads to the degradation of extracting semantic features.
Therefore, under the premise of avoiding non-negligible efficacy degradation, we achieve low computational complexity. The experiment results are shown in the next subsection.

\subsection{Scalability}

According to the results of efficacy evaluation, our method and SG-Lppm perform much better than others.
So, in this subsection, we compare the scalability of these top two methods in two aspects: Memory Consumption and Time Consumption.

\subsubsection{Memory Consumption}

We evaluate the memory consumption of our method and SG-Lppm. When establishing the generator of impostor traces, we store the necessary data in memory, which contain the semantic similarity of grids, and the users' mobility model. The space complexity of ours is $O(|R|^2+|R|\cdot{N_M})$, compared with $O(|R|^2+N\cdot|R|\cdot{N_U})$ for SG-Lppm, where $N$ is the number of input traces, and $|R|$ is the number of regions.

\begin{figure}
\centering
\includegraphics[width=0.4\textwidth]{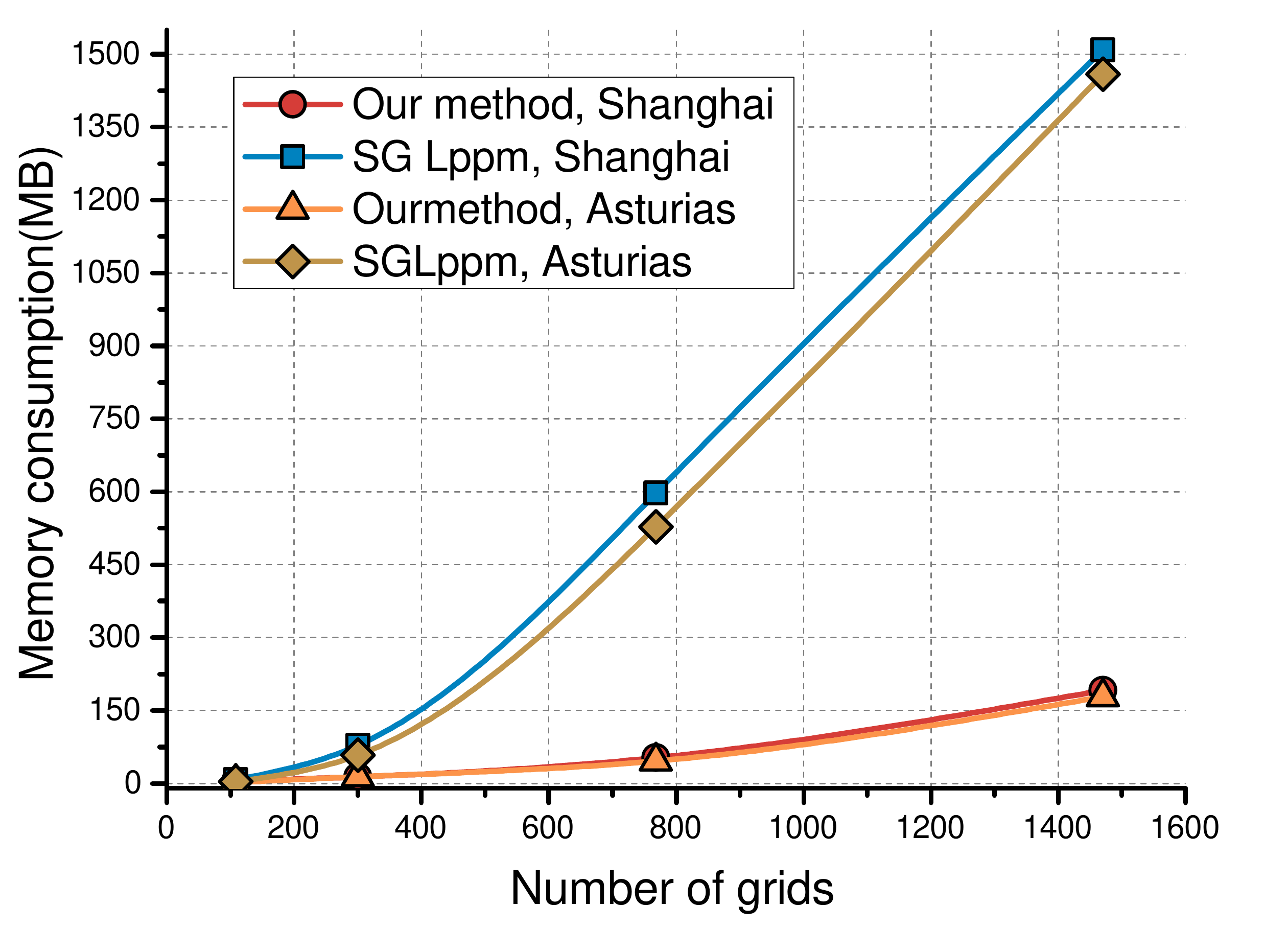}
\caption{Memory Consumption of Building the Module.}
\label{memorysdfdsfconsumption}
\end{figure}

In the experiment, we compute the memory consumption when using different scales of maps ($12\times9$, $20\times15$, $32\times24$, and $42\times35$ grids of area $1\mbox{km}\times1\mbox{km}$). We set $N_U=288$, and input $15000$ seed traces corresponding to different scales of maps to build generators.
Fig. \ref{memorysdfdsfconsumption} shows that the memory consumption and growth rate of ours are much less than SG-Lppm.
On the map of $42\times35$ grids in Shanghai city and Asturias, the consumptions of ours are $191.73$MB and $178.21$MB, respectively, while the consumptions of SG-Lppm are $1508.62$MB and $1459.07$MB, respectively.

\subsubsection{Execution Time}

\begin{figure}[t]
\centering
\subfigure[Dataset in Shanghai]
{\includegraphics[width=0.4\textwidth]{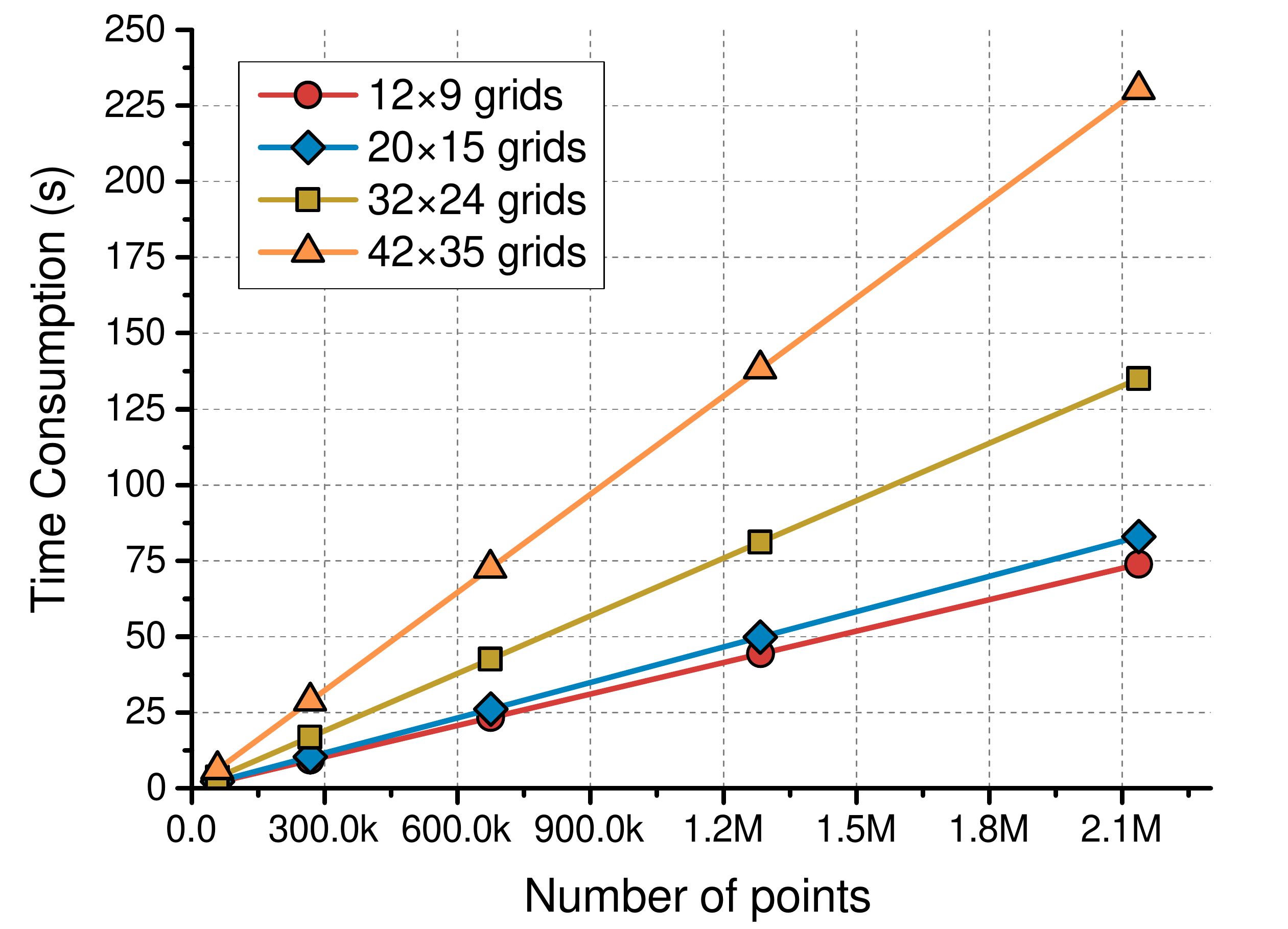}}
\subfigure[Dataset in Asturias]
{\includegraphics[width=0.4\textwidth]{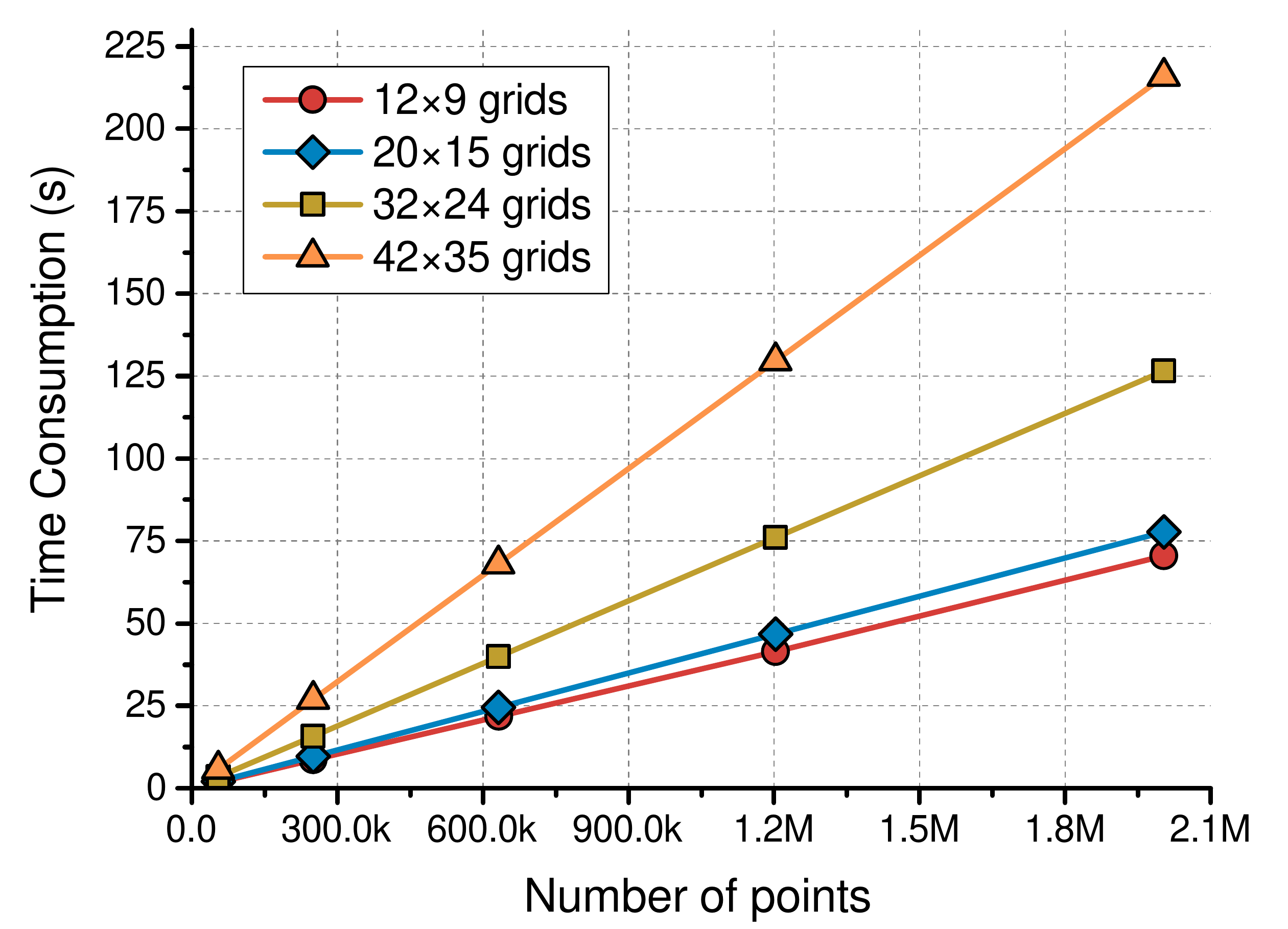}}
\caption{Runtime of Building the Module.}\label{sdfsdfscalability}
\vspace{-0.2in}
\end{figure}

We compute the runtime of building the offline generator between ours and SG-Lppm.
The time complexity of ours is $O(N\cdot{L}+|R|^3)$, compared with theirs
\[O(N^2\cdot|R|^3\cdot{N_U}+|R|^2\cdot{N_U}^2),\]
where $N$ is the number of seed traces (usually a large number), $|R|$ is the number of regions on the map, and $L$ is the average length of the traces ($L\leq{N_U}$).
It is straightforward that if we try to build the impostors generator,
it requires large amounts of input traces. In this evaluation,
we compare the execution time of $4$ scales of maps ($12\times9$, $20\times15$, $32\times24$, and $42\times35$ grids), and input $5$ different numbers of seed traces for each map when $N_U=288$.
The scales of the input data is the number of locations of the seed traces ($Num=N\cdot{L}$).
The relations between the scale of data and time consumption under different scales of maps are shown in Fig. \ref{sdfsdfscalability}.
On the map of $42\times35$ grids, the maximal scales $Num$ of the datasets in Shanghai and Asturias are about $2.1$ million and $2.0$ million, respectively. The time consumptions are $230.47$ seconds and $215.92$ seconds, respectively.
Under these conditions, we have executed SG-Lppm for two weeks on the server but without any output.
This is the reason why we only show the execution time of generating synthesizing model of our method in Fig. \ref{sdfsdfscalability}.
In contrast to such a frustrating result, while synthesizing $10$ impostors each time,
a preservation efficacy of $97.68\%$ and $96.24\%$ can be obtained by our method under this condition in two datasets, respectively.

\begin{table}[t]\renewcommand{\arraystretch}{1.6}
\centering
\caption{Runtime of generating one impostor trace.}\label{runtime}\centering
\begin{tabular}{p{1.6cm}|p{1.2cm}|p{1.2cm}|p{1.2cm}|p{1.4cm}}
\hline
\ Map Scales&$~~108$&$~~300$&$~~768$&$~~1470$\\
\hline
\ ~~Shanghai&$19.76$ms&$24.11$ms&$53.33$ms&$115.24$ms\\
\hline
\ ~~Asturias&$17.89$ms&$21.71$ms&$51.77$ms&$110.32$ms\\
\hline
\end{tabular}
\label{runtimetable}
\vspace{-0.1in}
\end{table}

Moreover, we test the time consumptions of generating traces through the online impostors' generators on different scales of maps: $12\times9$, $20\times15$, $32\times24$, and $42\times35$ grids (the size of grids are $1\mbox{km}\times1\mbox{km}$).
We record the time consumptions for synthesizing one impostor trace in different scales of maps in Table. \ref{runtimetable}.
SG-Lppm is really not scalable for large-scale maps and massive seed traces.
Therefore, we have to implement SG-Lppm on the map of $12\times9$ grids. The runtime of synthesizing one impostor trace is $72.31$ and $74.03$ seconds in the datasets of Shanghai and Asturias, respectively, much larger than ours $19.76$ms and $17.89$ms, respectively.

Recall that SG-Lppm \cite{synthisiziengsandp2016} analyzes the transfer probabilities among visited locations along traces of different users, and forges a fake location corresponding to every visited location when synthesizing fake traces.
Such refined method brings high preservation efficacy. However, such an individual-level approach causes impressively high time complexity.

In real-life applications, it is usually convincing to regard a user as a malicious one if the intervals between its adjacent queries are smaller than some small threshold. Thus, if the response time is smaller than such a threshold, then our method can work for any continuous queries from normal users.
Assuming that the threshold time is $s$ seconds,  the query rate of a normal user is then less than $1/s$ times per seconds.
From Table. \ref{runtimetable}, by our method, the response time for one query in different scales of maps is less than $0.116$ seconds.
We believe that $0.116$ seconds are sufficiently small compared with the threshold mentioned above.

\section{Conclusion}\label{sec-conclu}

We design a scalable and high-quality method for location privacy preservation based on the paradigm of synthesizing impostor traces.
Two dedicated techniques are devised:
the population-level semantic model and process-independent synthesis method.
Combining these two techniques, our method successfully achieves high preservation efficacy
with low computational
complexity.
Our method is proved to be capable of applying to the problems of different sizes.
We validate the scalability of our method from aspects of both memory consumption and execution time.

\ifCLASSOPTIONcaptionsoff
  \newpage
\fi

\clearpage

\end{document}